\documentclass[twocolumn,
preprintnumbers,prb,floatfix]{revtex4}

\usepackage{graphicx}
\usepackage{dcolumn}
\usepackage{bm}

\usepackage{amsmath,amssymb,amstext}
\usepackage[english]{babel}
\usepackage{textcomp}

\def\87Rb{$^{87}$Rb}
\def\mum{\text{\textmu m}}
\def\mus{\text{\textmu s}}
\def\muH{\text{\textmu H}}
\def\muW{\text{\textmu W}}
\def\muG{\text{\textmu G}}
\def\celsius{\text{\textcelsius} }

\newcommand{\ket}[1]{\ensuremath{\left| #1\right>}}
\newcommand{\bracket}[2]{\ensuremath{\left< #1\right|\left.#2\right>}}

\begin{document}

\preprint{Rev. Sci. Instrum.}

\title{Hybrid apparatus for Bose-Einstein condensation and cavity quantum electrodynamics: Single atom detection in quantum degenerate gases}

\author{Anton \"{O}ttl}
\author{Stephan Ritter}
\author{Michael K\"ohl}
\email{koehl@phys.ethz.ch}
\author{Tilman Esslinger}

\affiliation{Institute of Quantum Electronics, ETH Zurich, 8093
Zurich, Switzerland}

\date{\today}

\begin{abstract}
We present and characterize an experimental system in which we
achieve the integration of an ultrahigh finesse optical cavity with
a Bose-Einstein condensate (BEC). The conceptually novel design of
the apparatus for the production of BECs features nested vacuum
chambers and an \emph{in vacuo} magnetic transport configuration. It
grants large scale spatial access to the BEC for samples and probes
via a modular and exchangeable ``science platform.'' We are able to
produce \87Rb condensates of $5\times10^6$ atoms and to output
couple continuous atom lasers. The cavity is mounted on the science
platform on top of a vibration isolation system. The optical cavity
works in the strong coupling regime of cavity quantum
electrodynamics and serves as a quantum optical detector for single
atoms. This system enables us to study atom optics on a single
particle level and to further develop the field of quantum atom
optics. We describe the technological modules and the operation of
the combined BEC cavity apparatus. Its performance is characterized
by single atom detection measurements for thermal and quantum
degenerate atomic beams. The atom laser provides a fast and
controllable supply of atoms coupling with the cavity mode and
allows for an efficient study of atom field interactions in the
strong coupling regime. Moreover, the high detection efficiency for
quantum degenerate atoms distinguishes the cavity as a sensitive and
weakly invasive probe for cold atomic clouds.

\end{abstract}



\maketitle

%%%%%%%%%%%%%%%%%%%%%%%%%%%%%%%%%%%%%%%%%%%%%%%%%%%%%%%%%%%%%%%%%%%%%%%%%%%%%%%%%% introduction

\section{\label{sec:introduction}\textsc{Introduction}}
The research fields of Bose-Einstein condensation\cite{pethick2002}
(BEC) in dilute atomic gases and cavity quantum
electro\-dynamics\cite{berman1994} (QED) with single atoms both push
forward the understanding, engineering, and harnessing of quantum
mechanical states. A Bose-Einstein condensate is a collective
quantum state of a large atom sample and provides maximum control
over external degrees of freedom. Optical cavity QED in the strong
coupling regime allows probing and manipulation of single atoms with
the quantized electromagnetic field in the cavity mode.

A Bose-Einstein condensate is a fascinating demonstration of the
quantum character of matter where indistinguishable, weakly
interacting particles populate the motional ground state and
establish a macroscopic wave function. Its experimental
realization\cite{anderson1995,davis1995c} in 1995 sparked an ongoing
vivid experimental and theoretical research on this novel quantum
state. Initial experiments highlighted its phase
coherence,\cite{andrews1997}
superfluidity\cite{matthews1999a,raman1999} and demonstrated the
production of atom
lasers.\cite{mewes1997,anderson1998,hagley1999,bloch1999} Current
investigations explore quantum phase
transitions,\cite{greiner2002,stoferle2004} tunable atomic
interactions,\cite{inouye1998,donley2001} and particle
correlations.\cite{folling2005,ottl2005,schellekens2005}

\begin{figure}[b]
    \includegraphics[width=0.94\columnwidth, clip=true]{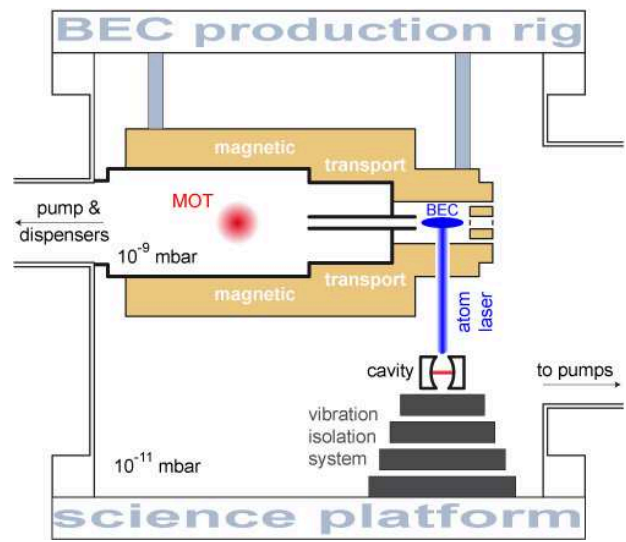}
    \caption{Schematic sketch of the experimental setup illustrating the nested vacuum chambers,
    the short magnetic transport and the ``science platform'' bearing the
    ultrahigh finesse optical cavity on top of the vibration isolation system.
    The atomic cloud captured in the magneto-optical trap (MOT) is transferred through a differential
    pumping tube into the ultrahigh vacuum region and evaporatively cooled towards quantum degeneracy.
    We output couple a continuous atom laser from the BEC and direct it to the cavity mode where single atoms are detected.}
    \label{fig:intro_schema}
\end{figure}

Similarly, the way to cavity QED in the optical domain was paved by
first experiments in the 1990's reaching the strong coupling regime
and demonstrating vacuum Rabi splitting of the coupled atom-cavity
system.\cite{thompson1992} In the strong coupling regime of cavity
QED the atom field interaction dominates over the dissipative losses
of the quantum system. An ultrahigh finesse cavity was used to
demonstrate single atom detection in an atomic
beam.\cite{mabuchi1996} Recent experimental progress was made in the
observation of the motional dynamics\cite{munstermann1999,hood2000}
as well as the trapping\cite{ye1999,pinkse2000} and
cooling\cite{van2001,maunz2004} of single atoms within the cavity
mode. This provides an avenue towards implementation of technologies
and concepts for quantum information processing, such as
nonclassical light sources\cite{kuhn2002,mckeever2004a} and quantum
state transfer.\cite{cirac1997}

The experimental combination of quantum degenerate gases with an
ultrahigh finesse optical cavity offers fascinating
prospects\cite{moore2000,jaksch2001,maschler2005} for quantum atom
optics. The first experiments detecting single atoms from a coherent
matter wave field with an ultrahigh finesse optical cavity have
recently been performed.\cite{ottl2005,bourdel2005} This progress
develops the new field of quantum atom optics where both matter and
light fields are quantized. Cavity QED detection of single atoms is
possibly nondestructive on the atomic quantum state and could be
used to perform atom interferometry with squeezed states and
precision measurements at the Heisenberg limit.\cite{bouyer1997} In
addition, the single atom detection method offers an unprecedented
sensitive and weakly invasive probe to investigate physical
processes in ultra cold atomic clouds \emph{in situ} and time
resolved. On the other hand, Bose-Einstein condensates and atom
lasers provide dense and coherent atomic sources with precisely
controlled atomic external degrees of freedom for exploring and
exploiting cavity mediated single atom single photon interactions.

The integration of an ultrahigh finesse optical cavity in a
Bose-Einstein condensation system, despite being a central goal for
atom chips,\cite{horak2003,eriksson2005} has only recently been
achieved\cite{ottl2005,bourdel2005} with the apparatus described
here. The experimental difficulties in merging these two
experimental research fields arise mainly from adverse vacuum
requisites and sophisticated topological requirements on both of
these state-of-the art technologies. For example, limited spatial
access prevents the inclusion of a high finesse optical cavity in
conventional Bose-Einstein condensation setups.

Our apparatus (Fig.\,\ref{fig:intro_schema}) overcomes the
experimental challenges of integrating an ultrahigh finesse optical
cavity into a Bose-Einstein condensation machine with a conceptually
novel design. It provides spacious access to the condensate for
divers samples and probes which are modularly integrable on our
science platform. This is rendered feasible by means of a nested
vacuum chamber design, a high vacuum (HV) enclosure inside the
ultrahigh vacuum (UHV) main chamber and a short \emph{in vacuo}
magnetic transport. Two distinct pressure regions are required since
the two common stages towards Bose-Einstein condensation, a
magneto-optical trap (MOT) for laser cooling and trapping a large
number of atoms and evaporative cooling, have conflicting
requirements on their vacuum environment. We utilize a short
magnetic transport\cite{greiner2001,lewandowski2003} to convey the
cloud of cold \87Rb atoms from the MOT to the main chamber, where we
perform evaporative cooling to quantum degeneracy. From the
Bose-Einstein condensate we output couple a continuous atom laser
and direct it into the cavity mode. The ultrahigh finesse optical
cavity is integrated on the so-called science platform and rests on
top of an UHV compatible vibration isolation system which is vital
for its stable operation. The cavity is located 36\,mm below the BEC
and enables us to detect single atoms from a quantum degenerate
source.

In the following, the modular experimental building blocks of our
hybrid BEC and cavity QED apparatus are presented in more detail.
Then we describe the operation and highlight the performance of our
\linebreak quantum atom optics experiment.

%%%%%%%%%%%%%%%%%%%%%%%%%%%%%%%%%%%%%%%%%%%%%%%%%%%%%%%%%%%%%%%%%%%%%%%%%%%%%%%%%%%%%%%%% vacuum

\section{\label{sec:vacuum}\textsc{Vacuum System: The Nested Chambers Concept}}

The vacuum system presented here consists of two nested steel
chambers where the higher pressure (HV) MOT chamber is situated
inside the lower pressure (UHV) main tank. The HV region houses the
alkali dispenser source. Both vacuum regions are pumped separately
and a differential pumping tube maintains a pressure ratio of
10$^{2}$. The setup grants multiple optical access for laser cooling
as well as for observation and manipulation of the resulting
Bose-Einstein condensate.

\begin{figure*}[t]
    \includegraphics[width=1\textwidth, clip=true]{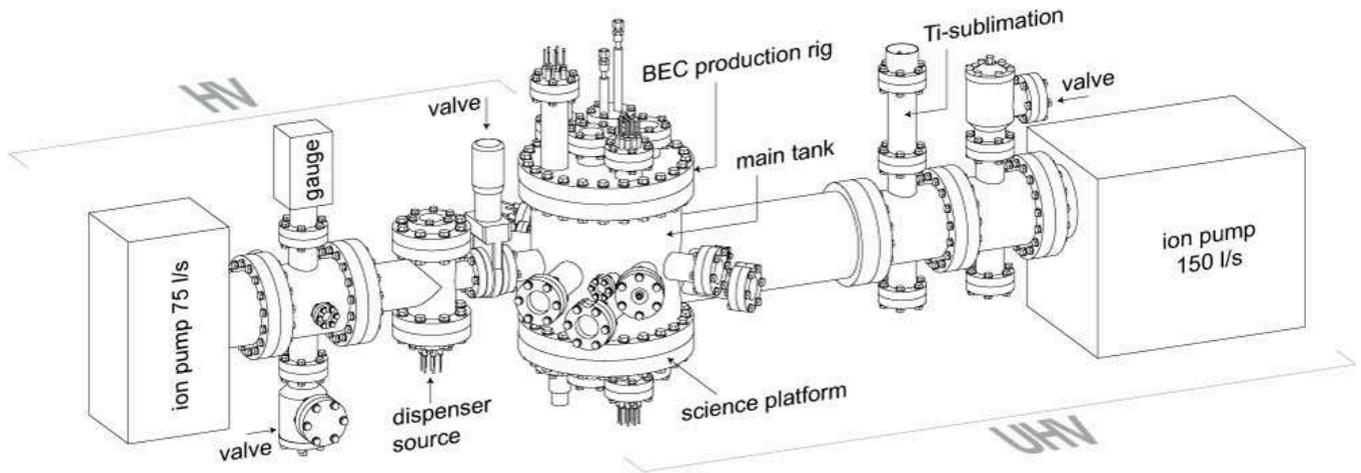}
    \caption{Overview of the complete vacuum system showing the pumping sections
    for the two nested vacuum regions, high vacuum (HV) and ultrahigh vacuum
    (UHV), respectively. The overall length is close to 2\,m.
    The main tank offers multiple optical and electrical
    access and is sealed off by two CF\,200 cluster flanges called ``BEC production rig'' and ``science platform.''}
    \label{fig:vac_all}
\end{figure*}

\subsection{\label{sec:vac_main}Main chamber}
The objective of the vacuum system (Fig.\,\ref{fig:vac_all}) is to
attain an UHV environment at $10^{-11}$\,mbar for efficient
evaporative cooling of the laser precooled atomic cloud. Centerpiece
of our vacuum system is the custom-welded, cylindrical main tank of
nonmagnetic stainless steel (AISI type 316).\cite{steel} It has a
diameter of 20\,cm and features multiple access (see
Fig.\,\ref{fig:vac_innen}) in form of optical grade viewports and
electrical feedthroughs with standard CF sealing. The viewports are
antireflection coated on both sides.

Two custom-made CF\,200 cluster flanges cap the main chamber from
above and below. The top flange (called ``BEC production rig'')
features optical and electrical access (Fig.\,\ref{fig:vac_innen})
since most of the electromagnetic coil configuration is mounted on
this flange and placed inside the UHV. In addition a liquid nitrogen
compatible feedthrough is supplied for cooling the magnet coils and
resistive temperature sensors (PT\,100) are used to monitor their
temperature. The bottom flange (called science platform) serves as
an exchangeable mount for the inclusion of samples and probes into
our system. Besides viewports and electrical feedthroughs
(Fig.\,\ref{fig:vac_innen}) to connect to electromagnetic coils,
PT100 sensors and the piezo element of the optical cavity design, it
includes a cold finger and a 300\,l/s nonevaporable getter (NEG)
vacuum pump.

The core vacuum pumping is performed by a titanium sublimation pump
and a 150\,l/s ion getter pump. A right angle valve is included in
this pumping section for rough pumping the system.

The HV part of the system (Fig.\,\ref{fig:vac_all}) connects to the
MOT chamber which protrudes into the UHV main chamber and serves as
a repository for rubidium atoms. It can be shut off with a gate
valve between the MOT chamber and the rubidium dispenser
source.\cite{wieman1995,fortagh1998} The HV region is pumped by an
ion getter pump (75\,l/s) whose pumping speed can be derated by a
rotatable disk inside the tube reducing its conductance. This serves
to control the rubidium vapor pressure which is monitored with a
wide range pressure gauge. Also a right angle valve is included for
rough pumping purposes.

Our rubidium repository consists of seven alkali metal dispensers
fixed in star shape to the tips of an eight pin molybdenum
electrical feedthrough where the center pin serves as the common
ground. Beforehand, the conductors were bent by $90\,^{\circ}$ so
that the dispensers aim towards the MOT chamber. Dispenser operation
may be viewed through a viewport mounted from above. The dispensers
can easily be exchanged without breaking the ultrahigh vacuum in the
main chamber by closing the gate valve between the MOT chamber and
the dispenser source.

\subsection{\label{sec:vac_mot}MOT chamber}
\begin{figure}[t]
    \includegraphics[width=1\columnwidth, clip=true]{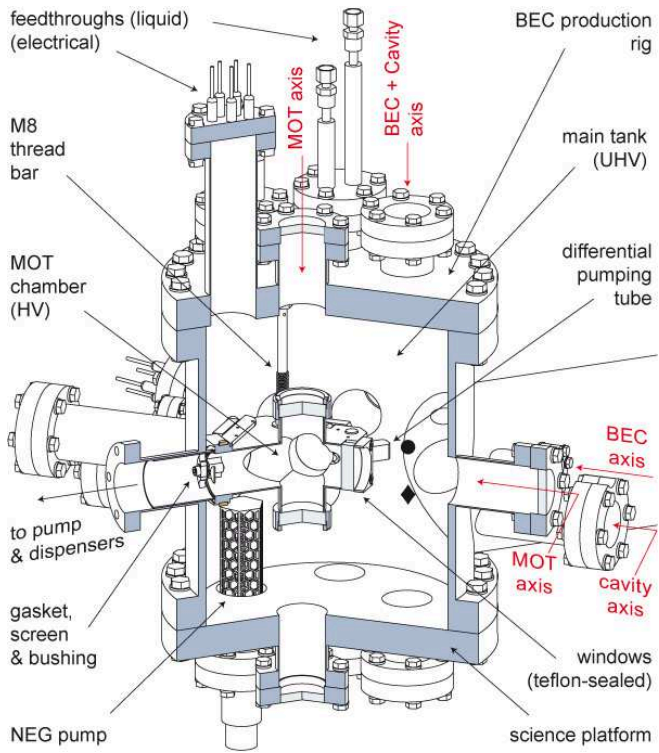}
    \caption{Section through the UHV system illustrating the realization of the nested chambers
    design and revealing the details and objectives of the divers optical axes.
    The position of the BEC and cavity are marked by
    ($\bullet$) and (${\scriptstyle \blacklozenge}$), respectively. The high vacuum MOT chamber is suspended from the ``BEC production rig''     and sealed by a tight fit bushing against the UHV main tank. The ``science platform'' provides space for
    additional components such as the ultrahigh finesse optical cavity.
    [Note: For clarity in the illustration the magnet coil configuration (Fig.\,\ref{fig:mag_coils-all})
    and the optical cavity assembly (Fig.\,\ref{fig:cav_platform}) are omitted in this
    figure.]
    }
    \label{fig:vac_innen}
\end{figure}

The MOT chamber as part of the high vacuum region is situated inside
the ultrahigh vacuum main tank (Fig.\,\ref{fig:vac_innen}). However
the fact that both pressure regions are well in the molecular flow
regime allows for relatively simple sealing techniques. The purpose
of the MOT chamber is to contain a higher background vapor pressure
of rubidium atoms for an efficient loading of the magneto-optical
trap.

Our MOT chamber was milled out of a single block of nonmagnetic
stainless steel (AISI type 316). This material was chosen to reduce
eddy currents produced by fast switching of the magnetic fields.
Bores of 35\,mm diameter give optical access for the six pairwise
counter-propagating laser beams forming the magneto-optical trap.
These bores are sealed off by standard optical grade laser windows
(BK\,7) with double-sided antireflection coating and clamped to the
MOT chamber by stainless steel brackets. At the metal glass
interface we use thin (0.2\,mm) Teflon rings to protect the windows.
Additionally we took precautions in the form of ceramic screens to
prevent coating of the windows by the titanium sublimation pump.

An additional bore provides the connection of the MOT chamber to the
HV pumping section and the dispenser source. This connection is
sealed against the UHV main tank with a tight fit stainless steel
bushing inside the CF\,40 socket. The bushing is tightened to the
MOT chamber thereby pressing its circular knife edge into a
custom-made annealed copper gasket. A screen to prevent a direct
line of sight from the hot dispensers to the center of the MOT is
included in the laser cut gasket.

The MOT chamber is sandwich mounted between the two magnet coil
brackets for the magnetic transport (see Sec.\,\ref{sec:magnetic})
and simultaneously functions as a spacer for the magnet coil
assembly. The whole structure is suspended from the top flange by
four M8 thread bars and represents our BEC production rig.

A differential pumping tube interfaces the MOT chamber with the main
tank. It serves for conveying the cloud of cold atoms with the
magnetic transport from the MOT into the UHV main chamber. The
aluminum differential pumping tube is mounted with a press fit in
the MOT chamber and can be exchanged. It has an inner diameter of
6\,mm over a length of 45\,mm and can maintain a differential
pressure of 10$^{2}$-$10^{3}$ depending on the actual pumping speed
in the UHV main chamber. Its conductance for rubidium at room
temperature is about 0.3\,l/s.

\subsection{\label{sec:vac_installation}Installation}
All components of the system were electropolished (the custom-welded
parts were pickled afore), cleaned and air baked at $200\,\celsius$
before assembly.\cite{cern1999} Additionally, all critical \emph{in
vacuo} materials like Stycast\,2850\,FT\cite{stycast} and
Kapton\cite{kapton} used for the magnet coil brackets (see
Sec.\,\ref{sec:magnetic}), Viton\,A\cite{viton} and
Wolfmet\cite{wolfmet} utilized for the vibration isolation stack
(see Sec.\,\ref{sec:cavity}) as well as plastic
(Teflon\cite{teflon}, Vespel\cite{vespel}) and ceramic
(Macor\cite{macor}, Shapal\cite{shapal}) parts were externally
outgassed by vacuum baking them at $200\,\celsius$.

The bakeout\cite{laurent1997} of the fully assembled system was
performed at $120\,\celsius$ which is the maximum temperature rating
of the piezotube used in our optical cavity assembly. The ultimate
attainable pressure in the UHV system is $3\times10^{-11}$\,mbar. It
is measured directly inside the main chamber in close proximity of
the magnetic trap for Bose-Einstein condensation. In the HV part we
maintain a pressure in the range of $10^{-9}$\,mbar.

%%%%%%%%%%%%%%%%%%%%%%%%%%%%%%%%%%%%%%%%%%%%%%%%%%%%%%%%%%%%%%%%%%%%%%%%%%%% magnetic

\section{\label{sec:magnetic}\textsc{Magnetic Field Configuration: \protect\\
Transport, Trap and Shielding}} A magnetic
transport\cite{greiner2001,lewandowski2003} is a reliable and
controlled way to transfer the cold atomic cloud from the MOT to a
region of considerably lower background pressure for evaporative
cooling. Only an \emph{in vacuo} magnet coil arrangement in
conjunction with nested vacuum domains allows for a short transport
design and grants spacious access volume inside the main chamber.
However care must be taken to meet the UHV requirements with the
materials chosen for the magnet coil structure.

Besides spatial and optical accessibilities the requirement on the
magnetic trap is mainly magnetic field stability to enable stable
atom laser output coupling. Therefore we employ a magnetic trap in
the quadrupole Ioffe configuration\cite{esslinger1998} (QUIC)
because its simplicity allows for a compact design and ensures an
easy and stable operation at very low power consumption
($\sim2$\,W).

A magnetic shielding enclosure and additional \emph{in vacuo} coils
for manipulating atoms in connection with the cavity round off the
magnetic configuration of the system.

\begin{figure}[b]
    \includegraphics[width=0.78\columnwidth, clip=true]{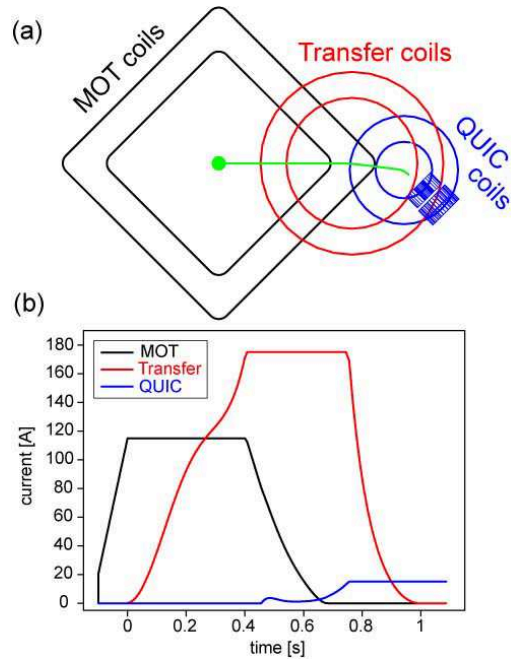}
    \caption{(a) Top view of the arrangement of coils for the magnetic transport.
    The line denotes the trajectory of the atomic cloud from the MOT (filled circle) into the
    QUIC trap. (b) Temporal sequence of currents
    through the different coils to realize the compression of the cold atomic cloud (negative times) and the magnetic transport.}
    \label{fig:mag_transport}
\end{figure}

\begin{figure*}[ht]
    \includegraphics[width=1\textwidth, clip=true]{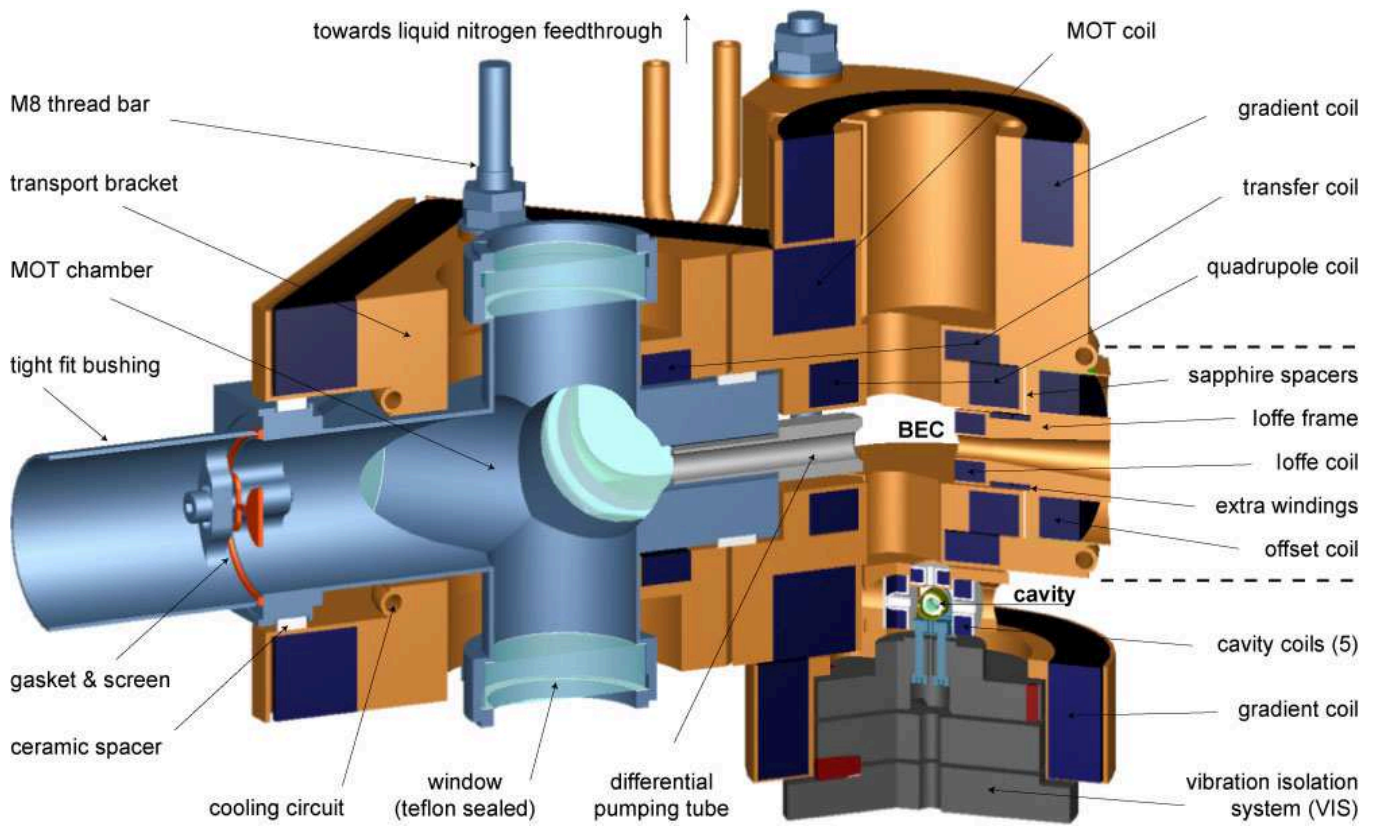}
    \caption{Section through the complete assembly
    inside the main vacuum chamber. It illustrates the arrangement of magnet
    coils, the inner chamber, and the cavity with respect to each
    other. Functional units of the magnet coil configuration are the
    two transport brackets that sandwich the inner chamber and the laterally mounted Ioffe
    frame (elements between dashed lines). These parts, including the top
    gradient coil, are fixed to each other and mounted from the top flange.
    The optical cavity on top of the vibration isolation
    system, the surrounding coils, and the bottom gradient coil are mounted on the science
    platform.}
    \label{fig:mag_coils-all}
\end{figure*}

\subsection{\label{sec:mag_transport}Magnetic transport} The magnetic
transport design consists of two partially overlapping
electromagnetic coil pairs (called ``MOT coils'' and ``transfer
coils'') producing quadrupole potentials and the final QUIC trap
coils [Fig.\,\ref{fig:mag_transport}(a)]. The overall potential
minimum can be moved over a distance of 82\,mm so that the cold
atoms in a low field seeking Zeeman state are conveyed from the
position of the MOT directly into the final magnetic QUIC trap. The
transfer coil pair provides sufficient overlap between the two to
achieve a smooth transfer of the magnetic potential without
significant heating of the cold atomic cloud.

The magnet coils were wound from rectangular copper wire
($3\!\times\!1$ and $1\!\times\!1$\,mm) for optimal filling
fraction. We choose Kapton film isolated wire which is temperature
durable and suitable in the ultrahigh vacuum environment. The coils
were integrated in two mirror-inverted, custom-made copper brackets
and encapsulated with Stycast\,2850\,FT, a thermally conductive
epoxy. The brackets are slotted in order to suppress eddy currents
from switching the magnetic field. The magnet coil assembly was
fixed in a sandwich structure around the MOT chamber and suspended
from the top flange by M8 thread bars
(Fig.\,\ref{fig:mag_coils-all}). The complete assembly including the
QUIC trap represents the BEC production rig.

A cooling system to remove the heat dissipated by the
electromagnetic coils is supplied in form of a copper pipe with
4\,mm inner diameter. It is soldered in a loop around each coil
bracket and connected to the liquid nitrogen feedthrough. A
temperature stabilized recirculating chiller\cite{chiller}
permanently pumps pure ethanol cooled to $-90\,\celsius$ through the
system. Thereby we maintain a maximum operating temperature below
$0\,\celsius$. This in turn lowers the power consumption. The
surface temperature of the coils is monitored with PT\,100 sensors
and interlocked to the power supplies.

The geometry and arrangement of the magnetic transport coils
[Fig.\,\ref{fig:mag_transport}(a)] are dominated mainly by
constraints set by the size of the MOT chamber, the required length
of the differential pumping tube, and the optical access to the MOT,
BEC, and cavity. For instance, the square shape of the MOT coils
best achieves a large overlap with the transfer coils while granting
optical access to the cavity axis. However, the aspect ratio $A/R$
of the coil separation $(2A)$ to the coil radius $(R)$ could be
tuned to a balanced tradeoff between a maximally strong $(A/R=0.5)$
and a maximally long $(A/R=0.87)$ linear gradient
region.\cite{bergeman1987} Anti-Helmholtz configuration is
advantageous for tight confinement and deep trap depths whereas long
linear gradients yield large handover regions between two coil
pairs. Furthermore, the power consumption of a coil pair for a given
field gradient can be minimized by choosing a well matched ratio of
axial to radial windings.

In order to find an optimum current sequence for the magnetic
transport\cite{greiner2001} we calculate the magnetic field of the
coil configuration analytically and discretize it along the
transport axis on a 100\,\mum\ grid. The currents needed to transfer
the magnetic minimum smoothly from the MOT to the QUIC are then
computed numerically in accordance with several constraints. Limited
by a maximum available electrical current we optimized the magnetic
field gradients and trap depths especially during the handover.
Furthermore we tried to minimize deformations of the trapping
potential. The resulting spatial sequence of currents per coil is
converted into a temporal sequence including an acceleration and
deceleration phase by taking into account the limited bandwidth of
the current control servo [Fig.\,\ref{fig:mag_transport}(b)].

The magnetic transport sequence initiates with a fast $(400\,\mus)$
ramp to 20\,A in the MOT coils after magneto-optical trapping and
optical pumping the cold atoms into a low field seeking state. The
ramp needs to be fast with respect to the expansion of the cloud but
adiabatic on the spin degree of freedom. It is followed by a slow
(100\,ms) compression of the atomic cloud to the maximum field
gradients. Increasing the current in the transfer coils pulls the
atoms towards their center and by decreasing the MOT coil current
the zero of the potential is handed over
[Fig.\,\ref{fig:mag_transport}(b)]. The field of the QUIC is aiding
at this point to maintain a constant aspect ratio. The magnetic
transport finishes by ramping down the current through the transfer
coils in favor of the QUIC coils. The atomic cloud is conveyed
through the differential pumping tube directly into the magnetic
QUIC trap which stays on for the subsequent evaporative cooling
stage.

The trajectory of the magnetic transport
[Fig.\,\ref{fig:mag_transport}(a)] is slightly bent such that atoms
in the final magnetic trap position have no direct line of sight
into the higher pressure MOT chamber. This suppresses background gas
collisions which would shorten the lifetime of the Bose-Einstein
condensate. The bend is achieved by laterally offsetting the center
of the QUIC trap by 3\,mm from the differential pumping tube.

The MOT and transfer coils are powered by general purpose interface
bus (GPIB) controllable 5\,kW dc power supplies.\cite{HPs} However,
since their internal current control bandwidth is too slow to sample
the time-current sequence for the MOT coils we externally feedback
control it by a closed-loop servo. It is implemented with a current
transducer and a MOSFET bench. The fast initial ramp to 20\,A is
additionally supported by current from four large capacitors (1\,mF)
charged to 60\,V. The electromagnetic properties of the coils with
resulting maximum currents and field gradients are listed in
Table\,\ref{mag:table}.

\begin{table}[t]
\caption{\label{mag:table} Electromagnetic properties of the magnet
coils.}
\begin{ruledtabular}
\begin{tabular}{lcccc}
& & MOT & Transfer & QUIC\\
\hline
Resistance & (m$\Omega$) &  200 & 50 & 300 \\
Inductance & (\muH) & 1000 &  70 & 450 \\
Maximum current & (A) & 115 & 170 & 15 \\
Maximum field gradient & (G/cm) & 310 & 290 & 320 \\
\end{tabular}
\end{ruledtabular}
\end{table}

The magnetic transport is performed over a period of 1\,s. We
maintain a minimum trap depth of $\sim 70$\,G equivalent to about
2\,mK. The total power required is approximately 2\,kW which
corresponds to an average power consumption of $\sim 34$\,W at a
duty cycle of 1/60.

\subsection{\label{sec:mag_quic}QUIC trap}
The magnetic QUIC trap consists of three coils connected in series.
This is advantageous to diminish relative current fluctuations and
therefore magnetic field fluctuations. Two coils (called
``quadrupole coils'') produce a quadrupole field and one smaller
coil (called ``Ioffe coil''), mounted orthogonally between the
quadrupole coils lifts the magnetic zero to a finite value and adds
a curvature to the resulting potential.\cite{esslinger1998} Having a
nonzero magnetic minimum is crucial when evaporatively cooling atoms
towards quantum degeneracy in order to circumvent losses due to
Majorana spin flips.

The geometry of the QUIC trap potential is approximately
cylindrically symmetric with respect to the Ioffe coil axis. Along
this direction the curvature and therefore the confinement is weaker
than in the radial directions. In our case this results in cigar
shaped Bose-Einstein condensates with an aspect ratio of 5:1.

The exact position and dimension of the Ioffe coil are very critical
to yield the desired magnetic bias field $B_{0}$ which should be on
the order of a few Gauss. A low bias field is preferential because
the trap frequencies scale as $B'/\sqrt{B_{0}}$, where $B'$ is the
magnetic field gradient and high trap frequencies permit faster and
more efficient evaporative cooling.

The construction of the Ioffe coil is done in the same way as for
the transport coils. It is integrated in a slotted copper frame and
potted with Stycast. The Ioffe frame is mounted laterally between
the transport coil brackets which hold the quadrupole coils.
Additionally, the Ioffe frame serves as a spacer for the two
transport brackets. The mechanical contact is accomplished with
sapphire sheets in order to prevent eddy currents by simultaneously
maintaining good thermal conductivity
(Fig.\,\ref{fig:mag_coils-all}). The large mass of the complete
magnet coil structure functions as a thermal low pass filter which
contributes to the good temperature stability.

In the Ioffe frame we have integrated additional coils on the same
axis as the Ioffe coil to be able to manipulate the final trap
geometry inside the vacuum system after bakeout. Two few-winding
coils are employed to fine-tune the value of the magnetic field
minimum $B_{0}$. One larger coil (called ``offset coil'') a little
further away from the trap center can be used to change the aspect
ratio of the trap and make it approximately spherical. Furthermore,
the Ioffe frame features a conical bore which allows us to image the
BEC through the center of the Ioffe coil.

The electrical connections of the coils forming the magnetic QUIC
trap are realized outside the vacuum. We have included a 1.4\,MHz
low pass filter in parallel to the Ioffe coil to avoid any radio
frequency (rf) pickup because of its low inductance of 4\,\muH. The
QUIC trap is operated with a 150\,W power supply\cite{power}
specifically tuned to our inductive load. The average power
consumption of the magnetic trap is maximally 60\,W but can be as
low as 2\,W when operated at 3\,A (see Sec.\,\ref{sec:operation}).

\subsection{\label{sec:mag_shielding}Magnetic shielding}
\begin{figure}[htb]
    \includegraphics[width=0.9\columnwidth, clip=true]{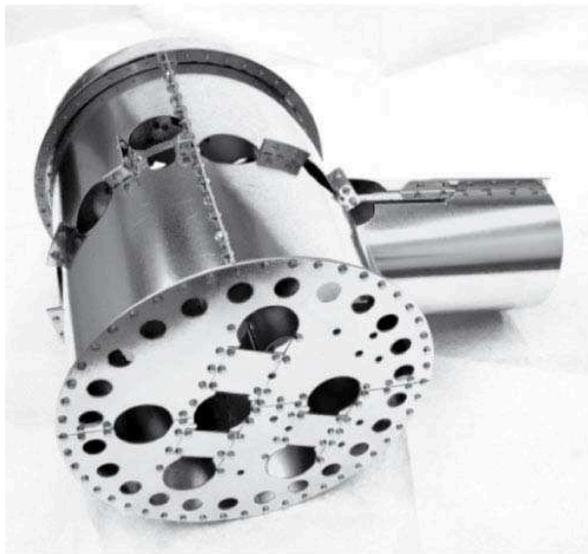}
    \caption{Photograph of the preassembled mu-metal hull before it is mounted around the main vacuum tank.
    It consists of seven large and several small individual pieces.}
    \label{fig:mag_shielding}
\end{figure}

We clad the main vacuum chamber in a mu-metal shielding
(Fig.\,\ref{fig:mag_shielding}) to minimize the influence of
residual external magnetic field fluctuations on the cold atoms. A
magnetically quiet environment is essential for stable continuous
wave (cw) operation of the atom laser.

Mu-metal is a magnetically soft nickel alloy with a very high
magnetic permeability $\mu \sim 10^{5}$ which attenuates magnetic
fields inside a cohesive enclosure. The screening effect depends
very much on the completeness of the mu-metal box. Magnetic field
lines penetrate an opening roughly as far as its diameter. Therefore
we have attached a stub around the pumping tube of the main vacuum
tank to attain a better aspect ratio at the position of the BEC. The
design of the mu-metal hull was aided by computer simulations of the
electromagnetic field. The mu-metal was machined and cured as
recommended by the manufacturer.\footnote{Vakuumschmelze} After
demagnetization we have measured a dc magnetic extinction ratio of
$\sim 40$ in the vertical and $\sim 100$ in the horizontal direction
at the position of the BEC.

\subsection{\label{sec:mag_auxiliary}Auxiliary coils}
Since the mu-metal shielding prevents any manipulation of the atoms
with external magnetic fields, we have arranged supplementary magnet
coils inside the mu-metal enclosure. All extra coils were potted
with Stycast either in a slotted copper or Shapal frame for good
thermal conductivity and mechanical sturdiness.

\begin{figure}[b]
    \includegraphics[width=1\columnwidth, clip=true]{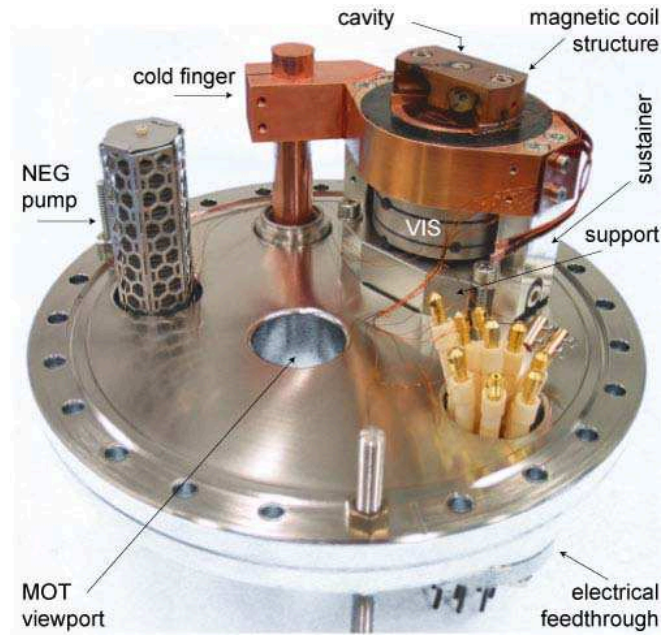}
    \caption{Photograph of the mounted science platform. The support bears the vibration isolation system (VIS)
    and the magnetic coil structure which surrounds the optical cavity.}
    \label{fig:cav_platform}
\end{figure}

Two large coils (called ``gradient coils'') are included in the main
vacuum chamber to compensate the gravitational force for the weakest
magnetic sublevel (30.5\,G/cm) with 22\,A. Their total resistance
and inductance is about 0.2\,$\Omega$ and 0.9\,mH, respectively. The
gradient coils were mounted inside the vacuum chamber on the
transport bracket (Fig.\,\ref{fig:mag_coils-all}) and on the science
platform around the cavity (Fig.\,\ref{fig:cav_platform}),
respectively. With the latter we should be able to reach the widest
Feshbach resonance of \87Rb ($\sim1008$\,G)\cite{marte2002} at the
position of the cavity.

Around the cavity we have placed two pairs of tiny coils
(4\,$\Omega$, 0.4\,mH) along and perpendicular to the cavity axis
(Fig.\,\ref{fig:mag_coils-all}). They can be used to create magnetic
field gradients of about 200\,G/cm (with 1\,A) for tomography
experiments. In combination with a fifth tiny coil (1\,$\Omega$,
0.1\,mH) mounted above a magnetic trap at the position of the cavity
can be formed. These five small coils (called ``cavity coils'') were
wound from 0.04\,mm$^{2}$ Kapton isolated copper wire on Shapal
frames to be penetrable by radio frequency.

In addition to the magnet coils inside the vacuum tank we have wound
three mutually orthogonal pairs of large extra coils around the main
tank. However, they are still within the mu-metal hull and serve to
produce homogeneous magnetic fields, e.g., for optical pumping.

%%%%%%%%%%%%%%%%%%%%%%%%%%%%%%%%%%%%%%%%%%%%%%%%%%%%%%%%%%%%%%%%%%%%%%%%%%%%%%%%% cavity

\section{\label{sec:cavity}\textsc{Science Platform: \protect\\
Implementation of the Optical Cavity}} We have designed this
apparatus with attention to versatile access for samples and probes
to the BEC. Therefore we have implemented two independent sections
of complementary functionality, i.e., the BEC production rig (see
Secs.\,\ref{sec:vacuum} and \ref{sec:magnetic}) and the science
platform. The latter is a modular, interchangeable flange, which in
the current configuration supports our single atom detector in form
of the ultrahigh finesse optical cavity.

The design of the cavity was guided by the need for stability,
compactness, and ultrahigh vacuum compatibility. It rests on top of
a passive vibration isolation stack which can be positioned on the
science platform.

\begin{figure*}[t]
    \includegraphics[width=1\textwidth, clip=true]{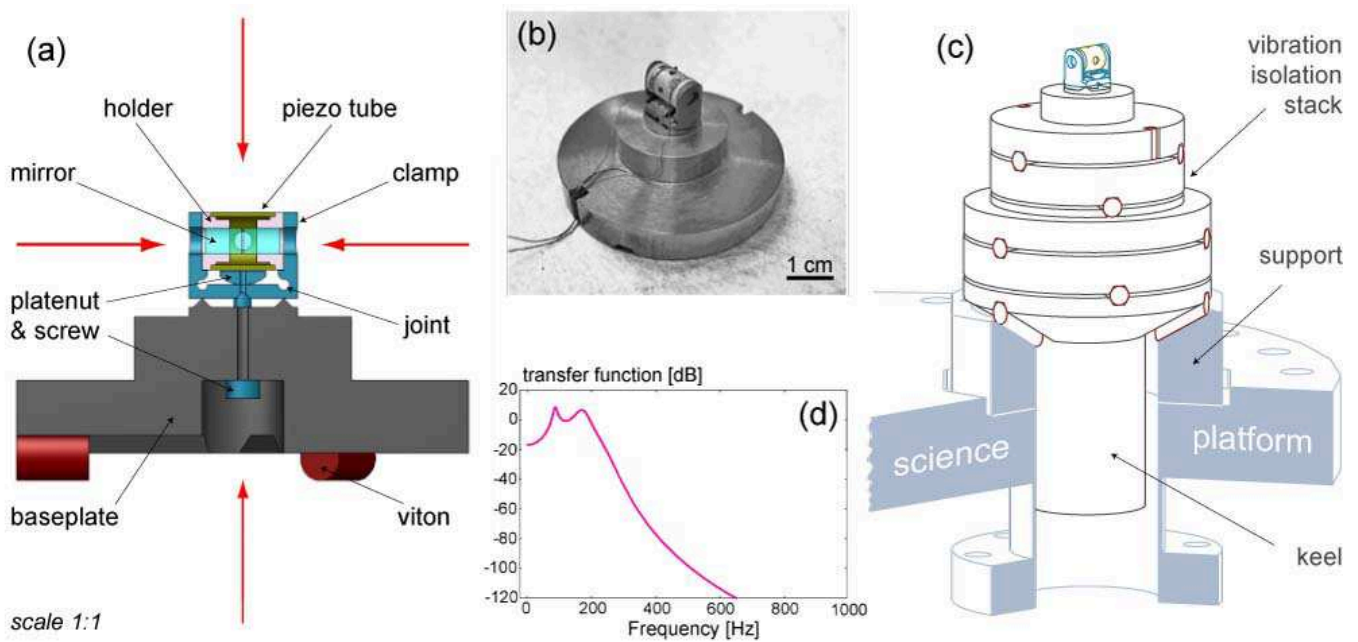}
    \caption{Elements of the optical cavity implementation.
    (a) Plane cut through the assembled cavity design where the arrows indicate optical
    access.
    (b) Photograph of the cavity setup. The electrical leads for the piezotube are pinched in a slotted Viton
    piece to efficiently decouple the cavity from the environment.
    (c) The cavity assembly resting on top of the vibration isolation stack which is positioned on the science platform.
    (d) Modeled frequency response of our vibration isolation stack.}
    \label{fig:cav_design}
\end{figure*}

\subsection{\label{sec:cav_design}Cavity design}
The Fabry-P\'erot optical cavity is formed by two dielectric Bragg
mirrors of ultrahigh reflectivity and ultralow scattering
losses.\cite{mirrors} The reflection band is 40\,nm wide and
centered around 780\,nm. We have determined an ultimate quality
factor $\text{Q}=1.6\times10^{8}$ after bakeout from the linewidth
of the cavity ($\Delta\nu = 2.4\,$MHz). The initial Q immediately
after cleaning the mirrors was higher by about a factor of two. The
cylindrical mirrors (3\,mm diameter, 4\,mm length) having a radius
of curvature of 77.5\,mm are separated by 178\,\mum which results in
a Gaussian mode waist of $\text{w}_{0} = 25.5\,$\mum. We precisely
measured the length of the near planar cavity by simultaneously
transmitting two different known wavelengths (see
Sec.\,\ref{sec:op_cavity-lock}) and determined a free spectral range
of $\nu_{\scriptscriptstyle \text{FSR}} = 0.84$\,THz from which we
derive a finesse of $\mathcal{F} = 3.5\times10^{5}$.

Each mirror was bonded with superglue into a specifically fabricated
ceramic (Shapal) ring structure. It positions and fixes the mirror
inside the piezoceramic tube.\cite{munstermann1999a} A piezo is
required to fine-tune the length of the cavity ($\sim0.5\,$V/nm) and
as the actuator for the cavity lock (see
Sec.\,\ref{sec:op_cavity-lock}). The 7\,mm long
piezotube\cite{piezo} has inner and outer diameters of 5.35 and
6.35\,mm, respectively. It is equipped with nonmagnetic wraparound
electrodes (silver) which allows the inner electrode to be contacted
from the outside. Additionally, the piezotube features four radial
holes of 1\,mm diameter for lateral access of atoms and lasers
perpendicular to the cavity axis.

The cavity assembly is mounted by a specifically designed compact
fixture (called ``clamp'') making use of mechanical joints
[Fig.\,\ref{fig:cav_design}(a)]. It was manufactured by spark
erosion from titanium in order to be nonmagnetic while having good
elastic properties. Further design considerations aimed at high
mechanical eigenfrequencies to avoid resonances within the bandwidth
of the cavity lock ($\sim 40\,$kHz), that means a small size and
high stiffness are favorable. We estimate the lowest eigenfrequency
of our fixture with a simple mechanical fixed-hinged beam
model\cite{harris2002} to be $\sim 50$\,kHz.

Our design of the cavity mount consists of the $\bm{\sqcup}$-shaped
clamp and a baseplate with integrated bearings to which the clamp is
tightened with a plate nut. It converts the downward force onto the
cavity assembly and firmly holds it together. Moreover it provides
the piezo with a load. A hole of 1.2\,mm diameter in the baseplate
and plate nut grants optical access to the cavity from below. This
cavity setup is highly modular and easily interchangeable because it
freely rests on the vibration isolation stack
[Fig.\,\ref{fig:cav_design}(b)].

\subsection{\label{sec:cav_vibration}Vibration isolation system} The
aforementioned baseplate simultaneously acts as the top mass of our
vibration isolation stack\cite{gerber1986} which consists of five
layers of massive plates (Wolfmet) with rubber dampers (Viton\,A) in
between [Fig.\,\ref{fig:cav_design}\,(c)]. Viton has good vibration
damping properties and is suitable for an ultrahigh vacuum
environment. The 5\,mm diameter Viton pieces rest in hexagonal
grooves that are radially arranged in $120\,^{\circ}$ graduations.
Consecutive layers are rotated by $60\,^{\circ}$ to prevent a direct
``line of sound.'' Hexagonal shaped grooves best avoid squeezing and
creeping of the rubber and provide good lateral stability. Position,
angle, and tilt reproducibility of this structure are excellent
because of the frustum shaped bottom mass with keel. It centers the
stack in an inverted, truncated conelike support and assures
mechanical stability by lowering the center of mass below the
support points. The complete stack has a central 10\,mm bore for
vertical optical access to the cavity.

Its damping properties can be modeled by regarding the structure as
a system of coupled masses and springs\cite{okano1987} and
calculating its frequency dependent transfer function
[Fig.\,\ref{fig:cav_design}(d)]. For attenuation at low frequencies
large masses and small spring constants are
favorable.\cite{oliva1992,oliva1998} Therefore we have fabricated
the plates from a heavy metal alloy (Wolfmet) and employed short
(10\,mm) Viton pieces. Our vibration isolation stack works well for
acoustic frequencies above 200\,Hz.

Additional precautions to counter low frequency excitations such as
building vibrations include setting up the experiment on a damped
rigid optical table in a basement laboratory having its own
independent foundation and choosing a position with little floor
vibration within this laboratory. The quality of the vibration
isolation system is such that we could easily operate the cavity in
the vicinity of a turbo-molecular pump. Furthermore the vibration
isolation stack kept the cavity in place when the whole optical
table accidentally dropped by about 2\,cm as we tried to tilt it.

\subsection{\label{sec:cav_platform} The science platform layout}
The self-contained, interchangeable science platform flange was
prepared to support and align the complete cavity mount. Its layout
provides manual positioning ability of the cavity mount by
\textpm\,2\,mm along and perpendicular to the cavity axis,
respectively. This is rendered feasible by an octagonal support
(nonmagnetic steel) of the vibration isolation stack which can be
deterministically moved and fixed in a larger octagonal millout on
the flange.

The second objective of the support is to erect the arrangement of
cavity coils with the gradient coil (see
Sec.\,\ref{sec:mag_auxiliary}) to be positioned around the cavity
without direct contact. The coil assembly is mounted on two
nonmagnetic steel sustainers which are fixed to the vibration
isolation support.

In order to remove the dissipated heat by the electromagnetic coils,
we have connected the copper bracket of the gradient coil to a power
feedthrough serving as a heat bridge, i.e., cold finger. Outside the
vacuum the 19\,mm diameter copper conductor can be connected to the
cooling circuit and cooled to $-90\,\celsius$. The copper rod serves
as a the main drain for the heat because of the low thermal
conductivity of the steel sustainers and support.

The mounting of the independent BEC production rig and science
platform within the main vacuum chamber has to be noncontact but
within a fraction of a millimeter. This results in a final position
of the optical ultrahigh finesse cavity being 36.4\,mm below the
BEC. The orientation of the cavity axis is at $90\,^{\circ}$ with
the symmetry axis of the magnetic trap (Ioffe axis).

%%%%%%%%%%%%%%%%%%%%%%%%%%%%%%%%%%%%%%%%%%%%%%%%%%%%%%%%%%%%%%%%%%%%%%%%%%%%%%%%% operation

\section{\label{sec:operation}\textsc{Operation: \protect\\
BEC, Atom Laser, and Cavity}} We operate the experiment periodically
with a cycle time of 60\,s. During each cycle we produce a new BEC
from which we output couple an atom laser. It is directed to the
ultrahigh finesse optical cavity situated 36.4\,mm below the BEC
where single atoms are detected. The cavity is probed by a resonant
laser and its length is actively stabilized by an off-resonant laser
with respect to the atomic transition.

The experimental sequence is fully computer controlled by a
\texttt{C++} program. Digital and analog channels interface the
computer with the elements of the experimental setup.\cite{cards}
The experiment is distributed on two self-contained optical tables,
one for the laser system and one for the vacuum apparatus. They are
linked by optical fibers.

\subsection{\label{sec:op_bec}BEC}
We form a Bose-Einstein condensate of \87Rb in dilute atomic vapor
from a dispenser loaded magneto-optical trap by means of rf-induced
evaporative cooling.\cite{ketterle1999,lewandowski2003}

During the first 20\,s of each cycle we load the magneto-optical
trap with atoms from the pulsed alkali dispenser
source.\cite{fortagh1998} The dispensers are operated at $\sim7\,$A
with a temporal offset of $-3\,$s to the actual MOT phase. We work
on the D2 line of \87Rb $(5\,^{2}\text{S}_{1/2} \rightarrow
5\,^{2}\text{P}_{3/2})$ at a wavelength $\lambda=780\,$nm. For the
cooling transition on the hyperfine ground state
$\ket{F=2}\leftrightarrow\ket{F'=3}$ a laser power of 17\,mW is
employed in each of the six 34\,mm diameter MOT beams. For optimum
collection efficiency we choose a detuning of $3\Gamma$, where
$\Gamma = 2\pi\,6$\,MHz is the linewidth of the cooling transition.
In order to be frequency tunable the laser is offset
locked\cite{schunemann1999} from the
$\ket{F=2}\rightarrow\ket{F'=2}$ transition by about 250\,MHz and
subsequently amplified with a tapered amplifier. An additional laser
(called ``repumper'') to avoid atomic losses into the \ket{F=1} dark
state is directly locked to the $\ket{F=1}\rightarrow\ket{F'=2}$
transition and delivers a power of 1\,mW in each MOT beam. All our
lasers are home built external cavity diode laser\cite{ricci1995}
locked by Doppler-free rf-spectroscopy technique\cite{bjorklund1983}
to atomic transitions. For the magneto-optical trap we apply a
magnetic field gradient of 10\,G/cm by applying a current of 3.5\,A
to the MOT coils. Due to the mu-metal shielding no earth field
compensation is required. We collect about $2\times10^{9}$ atoms
with the magneto-optical trap before we switch off the magnetic
field and sub-Doppler cool the atoms in a 10\,ms optical molasses
phase.

Before magnetically transporting the cold atomic cloud we optically
pump the atoms into the low field seeking \ket{F=1,m_{F}=-1}
hyperfine state. Optical pumping is performed over 2\,ms at a
homogeneous magnetic field of 4\,G. All light fields are off when
the transport sequence starts with adiabatically compressing the
cloud [Fig.\,\ref{fig:mag_transport}\,(b)]. The magnetic transport
conveys the atoms within 1\,s through the differential pumping tube
over a distance of 8\,cm directly into the magnetic QUIC trap. We
estimate a transport efficiency of $> 90\,\%$ by transferring the
atoms back into the magneto-optical trap and measuring their
fluorescence. The losses are mainly due to background collisions and
depend on the pressure in the MOT chamber.

We operate the magnetic QUIC trap initially with a maximum current
of 15\,A. This yields the highest trap frequencies of
$\omega_{x}=\omega_{z}=2\pi\,135$\,Hz and $\omega_{y}=2\pi\,28$\,Hz
with a bias field $B_{0}$ of 4.7\,G and a field gradient $B'$ of
$\sim300\,$G/cm. Here $\omega_{y}$ and $\omega_{x}$ denote the
trapping frequencies along and perpendicular to the Ioffe axis,
respectively and $\omega_{z}$ is in the vertical direction. Over a
period of 23\,s we perform rf-induced evaporative cooling with an
exponential frequency ramp and a radio frequency power of 24\,dBm.
The radio frequency is radiated by a coil which consists of ten
turns of Kapton clad copper wire (1\,mm$^2$) encircling an area of
3\,cm$^2$. It is mounted 2\,cm away from the center of the trap and
is oriented at $90\,^{\circ}$ with respect to the Ioffe axis. This
results in a $B_{\text{rf}}$ of about 30\,mG\ at the position of the
cold atoms.

Before reaching the critical phase space density for Bose-Einstein
condensation we relax the trap to the final parameters of
$\omega_{x}=2\pi\,38.5$\,Hz, $\omega_{y}=2\pi\,7.3$\,Hz and
$\omega_{z}=2\pi\,29.1$\,Hz with $B_{0}=1.2\,$G and $B'=60\,$G/cm by
powering the QUIC trap with 3\,A. The initial trap symmetry is
lifted by the large gravitational sag of about 290\,\mum. It is
given by $z_{\text{sag}}=-\text{g}/\omega_{z}^{2}$ where g is
Earth's gravitational acceleration. Furthermore, the long axis of
the BEC is inclined by about $20^{\circ}$ with respect to the
horizontal plane. The opening of the trap is performed adiabatically
($\dot{\omega} / \omega \ll \omega$) over a period of 1\,s. During
this time a rf shield limits the trap depth to prevent heating of
the cold atomic cloud. In the weak trap we further cool the atoms
evaporatively over 5\,s and achieve Bose-Einstein condensates of up
to $5\times10^{6}$ atoms. The density in the weak trap is
considerably lower so the losses due to inelastic collisions are
reduced. We have measured a 1/e-lifetime for condensates of about
30\,s. The typical size of the Bose-Einstein condensate is
$12\times15\times60\,\mum^3$ (Thomas-Fermi radii) with a chemical
potential $\mu$ of about 1\,kHz.

\begin{figure}[htb]
    \includegraphics[width=1\columnwidth, clip=true]{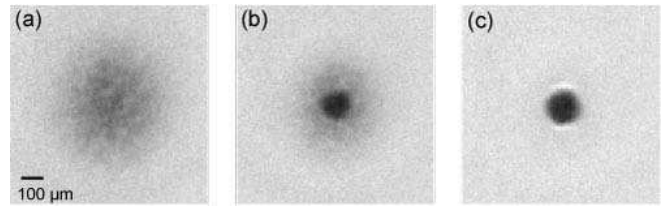}
    \caption{Absorption images of cold atom clouds.
    (a) Thermal cloud at a temperature $T$ above the critical temperature ${T}_{c}$.
    (b) Bimodal distribution for ${T}<{T}_{c}$.
    (c) ``Pure'' Bose-Einstein condensate at ${T}\ll{T}_{c}$.
    The images were taken after 30\,ms time of flight with a detuning of $2\Gamma$ to avoid saturation.}
    \label{fig:op_bec}
\end{figure}

Resonant absorption imaging of the cold atoms after a free expansion
time of 30\,ms allows us to extract the number of atoms in the cloud
and its temperature. We fit the resulting density distribution with
the sum of a Gaussian and a Thomas-Fermi profile. The spatial
resolution of our imaging system (f/10) is limited to 9\,\mum\ by
the diameter of the windows. We employ a charge coupled device (CCD)
camera with an according pixel size.\cite{apogee}

\subsection{\label{sec:op_atom-laser}Atom laser}
\begin{figure}[htb]
    \includegraphics[width=0.9\columnwidth, clip=true]{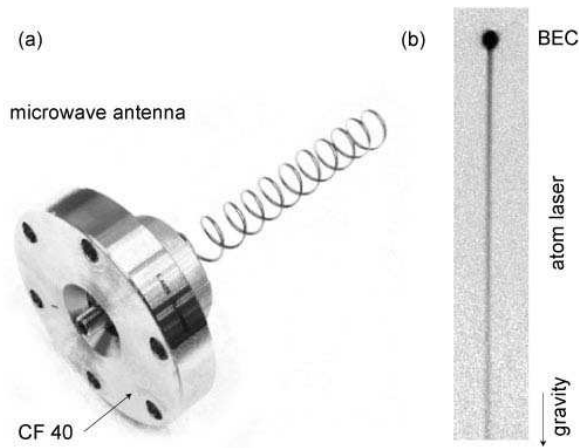}
    \caption{Coherent microwave output coupling of a continuous atom laser.
    (a) Helix antenna built for 6.8\,GHz mounted on a vacuum feedthrough.
    (b) Resonant absorption image of the atom laser after a propagation of 2\,mm.}
    \label{fig:op_atomlaser}
\end{figure}

An atom laser is a coherent atomic beam extracted from a
Bose-Einstein condensate (Fig.\,\ref{fig:op_atomlaser}). The trapped
condensate, being in a quantum degenerate state, serves as the
source for the freely propagating atom laser. A steady-state output
coupling process establishes a coupling between the ground state of
the trap and the energy eigenfunctions of the linear gravitational
potential and produces a continuous atom laser. The resulting cw
atom laser,\cite{bloch1999} in contrast to optical lasers, consists
of interacting massive particles propagating downwards in the
gravitational field. But like an optical laser it is a matter wave
in a coherent state as defined by Glauber in the quantum theory for
optical lasers\cite{glauber1963a} and exhibits higher order
coherence.\cite{ottl2005}

In order to output couple atoms we locally change their internal
spin state from the magnetically trapped \ket{F=1,m_{F}=-1} into the
untrapped \ket{F=2,m_{F}=0} hyperfine state. The spin flip is
induced by a coherent microwave field at the hyperfine splitting
frequency of \87Rb ($\Delta
E_{\text{hfs}}/h=6.8$\,GHz).\cite{steck2001} This microwave output
coupling scheme is equivalent to a two-level system because of the
Zeeman splitting in the hyperfine niveaus $(\sim 1\,$MHz). Therefore
it is superior to rf output coupling which mutually couples all
states from a Zeeman manifold.\cite{robins2005} The microwave signal
is produced by a global positioning system (GPS) disciplined
synthesizer.\cite{synthesizer} We use a home-built resonant helix
antenna with 14\,dB gain [Fig.\,\ref{fig:op_atomlaser}(a)] placed
inside the ultrahigh vacuum chamber to radiate the microwave field.
The antenna is connected and impedance matched to a commercial
microwave feedthrough.

The energy conservation for the microwave output coupling resonance
condition is only satisfied at regions of constant magnetic field
where $\Delta E_{\text{hfs}}-g_{F}m_{F}\mu_{\text{B}}B(\bm{r}) =
h\nu_{\text{mw}}$. Here $\nu_{\text{mw}}$ is the microwave
frequency, $B(\bm{r})$ the magnetic field of the trap at position
$\bm{r}$, and $\mu_{\text{B}}$ the Bohr magneton. The hyperfine
Land\'e $g$ factor $g_{F}$ and the magnetic spin quantum number
$m_{F}$ apply to the BEC state. The magnetic moment of the output
coupled atoms is zero to first order.

The resonant regions for output coupling are ellipsoidal shells with
the geometry of the magnetic trap, centered at the minimum of the
magnetic potential. However, the center of the actual harmonic
trapping potential for massive particles is lowered by the
gravitational sag with respect to the magnetic field minimum. For
our experimental conditions the gravitational sag is $\sim
290$\,\mum\ and therefore the resonant output coupling shells
intersect the Bose-Einstein condensate almost as horizontal planes.

The Rabi frequency $\Omega$ of the microwave output coupling process
is given by $\mu_{12}B_{\text{mw}}/\hbar$, where $\mu_{12}$ is the
magnetic dipole matrix element between the two coupled states and
$B_{\text{mw}}$ the magnetic field of the microwave
radiation.\cite{graham1999} The magnetic dipole transition has
selection rules $\Delta m_{F}=\pm1$. In the weak output coupling
regime ($\Omega\ll\omega_{z}$) an atom leaves the condensate much
faster than the Rabi frequency and does not undergo Rabi
oscillations.\cite{steck1998} The atom laser output coupling rate
depends on the number of atoms in the condensate $N_{\text{BEC}}$
and the overlap $|\bracket{\Psi_{\text{BEC}}}{\Phi_{\text{E}}}|^{2}$
between the BEC wave function $\Psi_{\text{BEC}}$ and the energy
eigenfunction $\Phi_{\text{E}}$ of the free atom
laser\cite{band1999,ballagh2000}. For given atom number
$N_{\text{BEC}}$ and microwave frequency the output coupling rate is
proportional to $\Omega^{2}$ and therefore to the power of the
incident microwave radiation\cite{schneider1999}.

Producing a coherent cw atom laser crucially depends on the temporal
stability of the resonance condition. We take experimental care to
avoid any fluctuations or drifts of the magnetic resonance position.
A temperature controlled cooling circuit for the large mass magnet
coil structure and a GPS locked synthesizer permit excellently
reproducible conditions. The magnetic shielding enclosure together
with the hermetic steel vacuum chamber eliminate external
electromagnetic field fluctuations (see Secs.\,\ref{sec:vac_main}
and \ref{sec:mag_shielding}). The only detectable noise source is
the low noise current supply powering the magnetic QUIC trap. We
have measured a magnetic field stability of better than
$5\muG/\sqrt{\text{Hz}}$ (at 3\,kHz) or 50\,\muG\ overall
(bandwidth: 50\,kHz), respectively. This enables us to produce
second order coherent atom lasers and output couple a cw atom laser
over the duration of the BEC lifetime. Due to the extremely low atom
fluxes measurable with the cavity detector we do not have to deplete
the condensate significantly.

The atom laser freely propagates downwards for 86.1\,ms before
entering the ultrahigh finesse optical cavity where single atoms are
detected. The cavity is placed 36.4\,mm below the BEC which results
in a velocity of 0.84\,m/s for the atoms traversing the cavity mode.
This velocity corresponds to a de\,Broglie wavelength of about 5\,nm
which could be useful for applications in coherent atom
lithography\cite{lee2000} or as an atom laser
microscope.\cite{balykin1987,doak1999}

\subsection{\label{sec:op_cavity-lock}Cavity lock}
In order to engage the ultrahigh finesse optical cavity as a single
atom detector we have to stabilize its length to better than
$0.5\,\lambda/\mathcal{F}\approx1\,$pm with respect to the
wavelength of the probe laser.

We choose a cavity locking scheme\cite{mabuchi1999} that allows us
to independently adjust the frequencies of the cavity resonance
$(\omega_{\text{c}})$ and of the probe laser $(\omega_{\text{l}})$.
Furthermore it enables us to keep the cavity permanently locked even
during atom detection since the action of a single atom transit on
the far-detuned stabilization laser is negligible and vice versa.

The cavity lock is realized with a far-detuned master laser at
830\,nm and a resonant master laser at 780\,nm referenced to a \87Rb
line. They are frequency stabilized by means of Pound-Drever-Hall
locks\cite{drever1983} to a transfer cavity having a free spectral
rage $\nu_{\scriptscriptstyle \text{FSR}}$ of 1\,GHz. In order to be
freely tunable the actual probe and stabilization slave lasers are
phase locked\cite{prevedelli1995} with a frequency offset of
0-500\,MHz to their respective master lasers. The length of the
science cavity is then actively controlled by a Pound-Drever-Hall
lock on the stabilization slave laser with a bandwidth of 38\,kHz.
We create the necessary sidebands for the lock with a home-built
electro-optical modulator.\cite{kelly1987} It works at 362\,MHz to
have the sidebands well off resonant with the cavity because its
finesse $\mathcal{F}$ for 830\,nm is only $3.8\times10^{4}$ and
therefore its linewidth $22$\,MHz.

We actively control the incident powers of the stabilization and
probe laser on the cavity to about 2\,\muW\ and 3\,pW, respectively.
In order to have a good spatial overlap, the two lasers are guided
through the same optical fiber. Their power ratio of $10^{-6}$ is
realized with an optical color filter. We can couple about 25\% of
the incident probe laser power into the cavity TEM$_{00}$ mode being
limited by the nonoptimal impedance matching.

The atomic resonance $(\omega_{\text{a}})$ we employ for single atom
detection is the cycling transition $\ket{F=2}\rightarrow\ket{F'=3}$
of the D2 line of \87Rb. It yields a maximum atom field coupling
rate
$g_{0}=d_{\text{iso}}\left|\bm{E}_{\text{max}}\right|=2\pi\,10.4$\,MHz,
where we have assumed an isotropic dipole matrix
element\cite{steck2001} $d_{\text{iso}}$ and a maximum single photon
electric field strength
$\left|\bm{E}_{\text{max}}\right|=\sqrt{4\hbar c/\left(\varepsilon_{
0}\lambda\text{w}_{0}^{2} l\right)}$ according to our mode volume
with a beam waist $\text{w}_{0}=25.5$\,\mum\ and a cavity length
$l=178$\,\mum.

The atom field coupling rate $g_{0}$ is large compared to the
dissipation losses being the cavity field decay rate
$\kappa=2\pi\,\Delta\nu/2$ and the dipole decay rate
$\gamma=\Gamma/2$ where $\Gamma=2\pi\,6.1\,$MHz is the natural
linewidth of the excited state. Furthermore the inverse atom transit
time $\tau^{\scriptscriptstyle -1}$ is orders of magnitude smaller
than the coupling rate which means the atom is always in a quasi
steady state with the cavity field during the transit. The relevant
parameters of our experiment are thus
$(g_{0},\;\gamma,\;\kappa,\;\tau^{\scriptscriptstyle
-1})=2\pi\,(10.4,\;3.0,\;1.2,\;3\!\times\!10^{-3})$\,MHz, which
brings us into the strong coupling regime of cavity QED defined by
$g_{0}\gg(\gamma,\;\kappa,\;\tau^{\scriptscriptstyle -1})$.

%%%%%%%%%%%%%%%%%%%%%%%%%%%%%%%%%%%%%%%%%%%%%%%%%%%%%%%%%%%%%%%%%%%%%%%%%%%%%%% detection

\section{\label{sec:detection}\textsc{Single Atom Detection}}
The single atom detection with an ultrahigh finesse optical
cavity\cite{mabuchi1996} can heuristically be viewed as the
refractive index of a single atom being sufficient to significantly
shift the cavity resonance. Consequently, the transmission of an
initially resonant, weak probe laser is measurably reduced. In
quantum mechanical terms the coupling of a single atom with the
quantized electromagnetic field in the cavity mode dominates the
dissipation losses (strong coupling regime) which means the level
splitting of the Jaynes-Cummings model\cite{jaynes1963,shore1993}
can be resolved. On the other hand, the quantum mechanical detection
process on the longitudinally delocalized atom within the atom laser
beam projects and localizes them inside the cavity
mode.\cite{bourdel2005}

We can efficiently study these cavity QED interactions of single
atoms having an atom laser as an unprecedented bright, controllable,
reproducible, and well defined atom source. Here we present
experimental results that characterize the performance of our
combined BEC and ultrahigh finesse optical cavity system.

\subsection{\label{sec:det_analysis} Analysis}
\begin{figure}[htb]
    \includegraphics[width=1\columnwidth, clip=true]{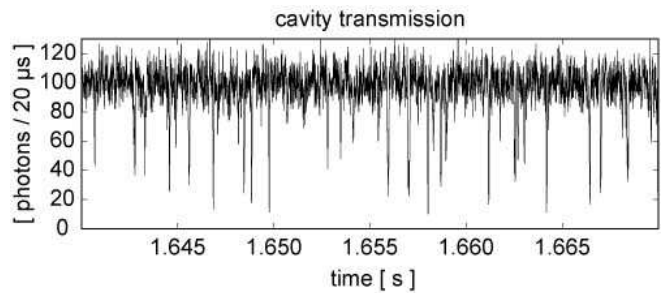}
    \caption{Cavity detection recording of an atom laser. The atom
    flux is about four orders of magnitude lower compared to
    Fig.\,\ref{fig:op_atomlaser}(b).
    Single atom transits are clearly identified by their reduction of the shot noise limited empty cavity transmission.}
    \label{fig:det_trace}
\end{figure}

In order to identify single atom transits we record the transmission
of a resonant weak probe beam through the cavity with a single
photon counting module\cite{SPCM} (SPCM). A typical recording
showing single atom transits is presented in
Fig.\,\ref{fig:det_trace}.

The light coming from the cavity is filtered with a 780\,nm band
pass and a 830\,nm notch filter to block the stabilization laser.
Their combined relative optical density (OD) for 830\,nm is 12. The
SPCM is located inside a blackbox and exhibits an overall photon
dark count rate of $\sim 100\,\text{s}^{-1}$.

The cumulative detection probability for intracavity probe photons
taking into account losses in the optical system and the quantum
efficiency of the SPCM is about 7\%. It is mainly limited by the
fact that we employ symmetric cavity mirrors with equal
transmittivity $(\sim2\,\text{ppm})$ and by the scattering losses
$(\sim7\,\text{ppm})$ of the mirrors.

In order to achieve a large signal-to-noise ratio for single atom
detection we usually work with an average intracavity probe photon
number of about 5.\cite{bourdel2005} This level corresponds to an
intensity of about 40 times the saturation intensity and yields a
photon count rate of $2\pi\,\Delta\nu\times5\times7\%\approx
5$\,photons/\mus.

We integrate the signal from the SPCM over 20\,\mus\ with a temporal
resolution between 1-4\,\mus\ and set the criterion for single atom
detection events to a reduction of more than four times the standard
deviation ($\sigma$) of the shot noise limited empty cavity
transmission. This reduces false atom detection events to less than
$0.5\,\text{s}^{-1}$.

\subsection{\label{sec:det_characteristics}Characteristics of single atom events}
\begin{figure}[htb]
     \includegraphics[width=1\columnwidth, clip=true]{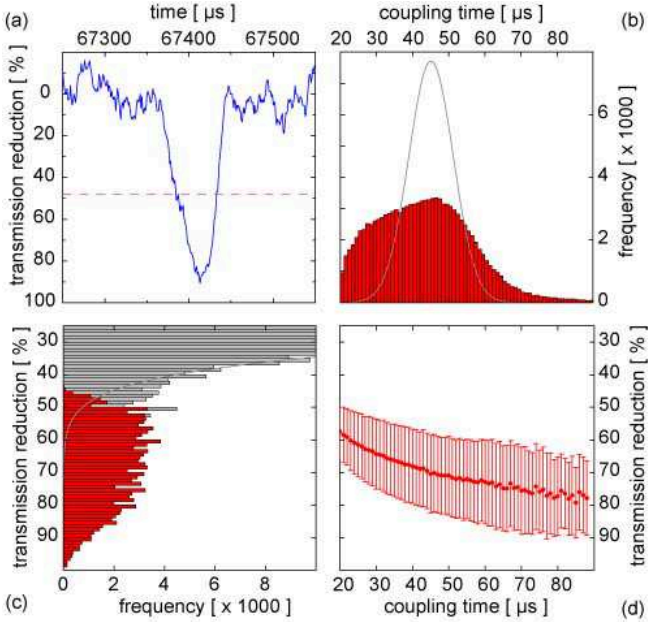}
    \caption{Characteristics of detected single atom events.
    (a) The transit of a single atom significantly reduces the probe light transmission through the cavity.
    We integrate the signal with a 20\,\mus\ sliding average and set the detection threshold to $4\sigma$ of the photon
    shot noise. (b) Distribution of measured coupling times (FWHM) [red] compared to the distribution of simulated events (gray). (c)
    Distribution of measured transmission reduction magnitudes. An evaluation
    with a $4\sigma$ threshold (red) is compared to a $2\sigma$
    threshold [gray] revealing the discrimination of the events from the photon shot noise. (d) Dependency
    of the transmission reduction on the coupling time due to the non-Gaussian shape of the dips.}
    \label{fig:det_characteristics}
\end{figure}

The coupling of a single atom with the cavity mode can be
characterized by the magnitude and duration of the resulting
transmission dips. A recorded typical single atom transit is shown
in Fig.\,\ref{fig:det_characteristics}(a). The $4\sigma$ threshold
here corresponds to about 50\% reduction in the probe light
transmission of about 70\,photons/\mus. We analyze detected events
and show histograms in Figs.\,\ref{fig:det_characteristics}(b) and
\ref{fig:det_characteristics}(c) for atom laser data taken in 184
iterations of the experiment. The atom flux was set to
$\sim1\times10^{3}\,s^{-1}$ so the probability\cite{ottl2005} for
unresolved multiatom events within the dead time of our detector
($\sim 70$\,\mus) is less than $0.3\%$.

The dead time of our detector is reflected in the distribution of
coupling times, i.e., the full width half maximum (FWHM) of the
transmission dips [Fig.\,\ref{fig:det_characteristics}(b)]. It is
mainly determined by the radial size of the Gaussian cavity mode and
the velocity of the atoms during their transit. For a radial
coupling strength
$\text{g}(r)=\text{g}_{0}e^{-{r^{2}}/{{\text{w}_{0}}^{2}}}$ with
w$_{0} = 25.5$\,\mum\ and an initial velocity of 84.1\,cm/s we
expect an average coupling time of $45\pm12$\,\mus\
[Fig.\,\ref{fig:det_characteristics}(b), gray]. Taking the classical
free fall velocity is justified since the induced momentum
uncertainty by projecting the longitudinally delocalized atom into
the cavity mode is on the order of 10\,\mum/s.

In the numerical simulation we take into account photon shot noise
and the features of our peak detect routine, namely, the 20\,\mus\
sliding average. The effect of the dipole potential on the transit
time is negligible because the slight gain in velocity
$(<\!2\,\mus)$ is counteracted by an effectively stronger and
therefore longer coupling [Fig.\,\ref{fig:det_characteristics}(d)].
The mean of the measured coupling time distribution
[Fig.\,\ref{fig:det_characteristics}(b), red] is in accordance with
the expected value. However, the distribution deviates from the
expected shape and exhibits an excess of short and long transit
times. We attribute the shorter transits to optical pumping of atoms
into the dark state \ket{F=1} because their number is intensity
dependent on the probe light. Longer transit times could be
explained by diffraction of the atomic beam, scattering of
spontaneous photons or cavity cooling effects, if the cavity axis is
slightly nonorthogonal with respect to the atom laser (possibly
$10^{-2}$\,rad) and by unresolved multiatom events.

The magnitudes of the cavity transmission dips
[Fig.\,\ref{fig:det_characteristics}(c)] reflect the different
maximum coupling strengths for single atom transits. Depending on
its radial position an atom will experience a varying peak coupling
strength according to the Gaussian profile of the cavity mode. In
the axial direction, however, the light force is strong enough to
channel the atoms towards the intensity maxima of the standing
wave.\cite{salomon1987} Arbitrarily weak coupling transits cannot be
resolved due to the shot noise in the empty cavity transmission. We
set the single atom detection threshold to $4\sigma$ of the original
transmission to achieve a large signal-to-noise ratio.

The resulting histogram of peak depths is displayed in
Fig.\,\ref{fig:det_characteristics}(c)(red) compared to data for a
lower threshold level of $2\sigma$ in
Fig.\,\ref{fig:det_characteristics}(c)(gray) unveiling the photon
shot noise. The weakest detectable single atom events correspond to
peak atom field coupling strengths of $\text{g}_0^{\text{min}} =
2\pi\,6.5$\,MHz. The strongest attainable coupling strengths for our
cavity are $\text{g}_0^{\text{max}} = 2\pi\,10.4$\,MHz, which would
be equivalent to a reduction of 80\% in the cavity
transmission.\cite{bourdel2005} We do not observe a sharp cutoff in
the histogram but rather an equal distribution of transmission
reductions from 50-80\% with smeared out edges due to the
comparatively large photon shot noise at the minimum of the
transmission dip. This is consistent with numerical simulations for
single atom events.

The dependence of remaining probe light transmission through the
coupled atom-cavity system is nonlinear with the atom field coupling
strengths.\cite{bourdel2005} Therefore the shape of the transmission
dips is not Gaussian as the cavity mode and we observe a dependency
of the magnitude in transmission reduction on the coupling time and
vice versa [Fig.\,\ref{fig:det_characteristics}(d)].

The knowledge about the signatures of single atom events could
facilitate the discrimination of ``true'' single atom events from
``false'' shot noise events or unresolved multiatom events, but the
broad distributions make it difficult to distinguish two weakly
coupling atoms from a strongly coupling one. However, the observed
characteristics of the detected events are in good agreement with
the theoretical predictions for single atom transits. Moreover,
these characteristics remain valid even when reducing the atom flux
to very few single atom events.

\subsection{\label{sec:det_detector}Detector qualities}
Having a BEC and an atom laser as the source for atoms that couple
with the cavity mode offers several advantages. For instance it
provides well reproducible starting conditions and allows us to
precisely control the flux of atoms over a wide range by varying the
microwave output coupling power. The attainable atom flux is orders
of magnitude larger than in experiments employing a magneto-optical
trap as the cold atom source.

\begin{figure}[htb]
    \includegraphics[width=0.8\columnwidth, clip=true]{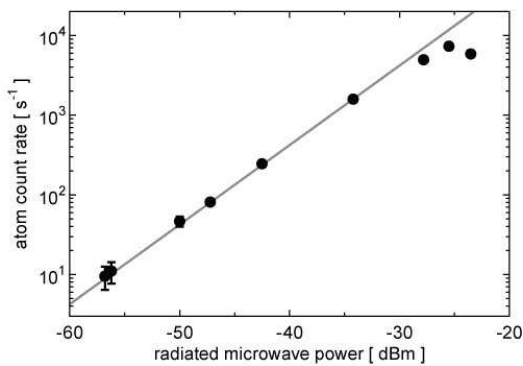}
    \caption{The ultrahigh finesse optical cavity functions as a linear detector
    on the output coupling rate, i.e., atom flux over three orders of magnitude.
    Saturation occurs at a count rate of about $5\times10^{3}$ atoms per second.}
    \label{fig:det_detector}
\end{figure}

We have confirmed that our single atom detector functions as a
linear detector on the atom flux over three orders of magnitude
(Fig.\,\ref{fig:det_detector}). The measured atom count rate is
proportional to the output coupling microwave power (see
Sec.\,\ref{sec:op_atom-laser}). Saturation occurs at a flux of about
$5\times10^{3}$ atoms per second. At higher rates multiatom arrivals
within the dead time of our detector become dominant and single atom
events cannot be resolved anymore.

At a very low atom flux the error bars become increasingly large due
to atom shot noise, i.e., the Poissonian distribution in the atom
number determination. Additionally, a very weak atom ``dark count''
rate without intentional output coupling may be present. It is
likely due to stray magnetic or optical fields and depends on the
size of the Bose-Einstein condensate. However, the dark count rate
is still less than 5 atoms per second on average for a BEC with
$2\times10^6$ atoms, for instance.

\subsection{\label{sec:det_efficiency}Detection efficiency}
The single atom detection efficiency of the ultrahigh finesse
optical cavity strongly depends on the frequencies
chosen\cite{pinkse2000a} for the probe laser ($\omega_{l}$) and the
cavity resonance ($\omega_{c}$) with respect to the atomic
transition ($\omega_{a}$). Furthermore the effective coupling
strength $\text{g}_{0}$ and therefore the detection probability are
determined by the polarization of the probe light with respect to
the quantization axis of the atomic spin.

In our experimental configuration we have a residual vertical
magnetic field at the position of the cavity of about 16\,G which
represents the quantization axis for the atoms. The field originates
from the magnetic QUIC trap which is on during the single atom
detection in the atom laser.

We set the probe light to horizontal (within $10\,^{\circ}$)
polarization which yields a four times higher atom count rate as
vertically (within $10\,^{\circ}$) polarized light. Only these two
distinct polarization settings are feasible since we experience a
birefringence in the cavity resonance of about twice its linewidth.
The horizontal polarization of the probe light produces a higher
atom field coupling rate because it drives $\sigma^{+}$ and
$\sigma^{-}$ transitions compared to the fewer and weaker $\pi$
transitions for vertically polarized light. The exact atom field
interactions are more complex because of the Zeeman splitting and
the resulting optical pumping dynamics inside the cavity.

However, for red detuned probe light the atoms entering the cavity
in the $\ket{F=2,m_{F}=0}$ state will predominantly be pumped into
the $\ket{F=2,m_{F}=-2}$ stretched state and undergo cycling
transitions driven by the $\sigma^{-}$ polarization component.
Therefore this cycling transition will be the main contribution in
the single atom detection process. The imbalance is due to a
redshift for the $\sigma^{-}$ component and a blueshift for the
$\sigma^{+}$ component of $\sim 22\,$MHz in the magnetic field of
16\,G at the cavity.

\begin{figure}[b]
    \includegraphics[width=1\columnwidth, clip=true]{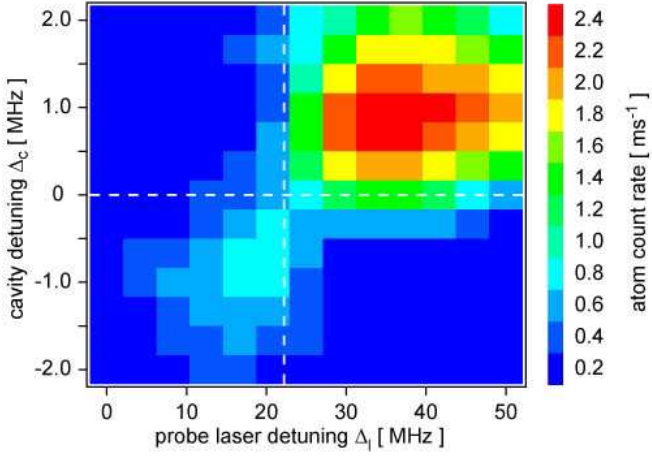}
    \caption{Dependence of the single atom detection efficiency on the probe laser $\Delta_{l}$ and cavity $\Delta_{c}$
    detunings. The vertical dashed line represents the cycling
    transition which is Zeeman shifted by 22\,MHz from the zero field
    atomic transition. Best single atom detection is performed with a probe laser
    red detuned by about $3\Gamma$ from the cycling transition and
    a cavity detuning of about $\Delta\nu/2$, corresponding to the
    maximum dipole potential created by the probe laser. The second local detection maximum corresponds to a blue detuned probe
    laser. Therefore the dipole potential is repulsive and the atom count rate reduced.}
    \label{fig:det_efficiency}
\end{figure}

The number of detected atoms critically depends on the atom-probe
laser detuning $\Delta_{l}=(\omega_{a}-\omega_{l})/2\pi$ and probe
laser-cavity detuning $\Delta_{c}=(\omega_{l}-\omega_{c})/2\pi$ as
illustrated in Fig.\,\ref{fig:det_efficiency}. Here $\omega_{a}$
refers to the bare atomic transition without magnetic field. For
most efficient single atom detection we work with an atom - probe
laser detuning $\Delta_{l}\approx$ 30-40\,MHz and a probe laser -
cavity detuning $\Delta_{c}\approx$ 0.5-1\,MHz. By taking into
account the 22\,MHz Zeeman shift of the cycling transition
$\ket{F=2,m_{F}=-2}\leftrightarrow\ket{F'=3,m_{F'}=-3}$ (vertical
dashed line in Fig.\,\ref{fig:det_efficiency}) the probe laser red
detuning for optimum single atom detection is about $3\Gamma$. This
value corresponds to the maximum of the dipole potential created by
the probe laser, that means the dipole force channels the atoms in
the axial direction towards the antinodes of the standing
wave\cite{salomon1987} which are simultaneously the areas of the
highest atom field coupling strength. In the radial direction the
dipole force is too weak to significantly modify the trajectory of
the atoms within the cavity mode. Also the dipole potential created
by the stabilization laser is weak compared to the one created by
the probe laser.

The second set of parameters in Fig.\,\ref{fig:det_efficiency} where
single atom transits are detected is around $\Delta_{l}\approx$
18\,MHz and $\Delta_{c}\approx-1$\,MHz. However, the count rate is
reduced considerably because the probe light is blue detuned from
the cycling transition and therefore the dipole potential is
repulsive. In the other two quadrants spanned by the resonances of
the cavity and the cycling transition of
Fig.\,\ref{fig:det_efficiency} (dashed lines), atom transits result
in increased probe laser transmission versus the empty cavity
transmission.\cite{pinkse2000} We do not use those events for single
atom detection because the efficiency is reduced by about a factor
of 2 as compared to evaluating dips. Additionally the peaks exhibit
a substructure consisting of single photon bursts which makes it
more difficult to discriminate single consecutive atom transits.

In order to determine the detection efficiency for single atoms from
the Bose-Einstein condensate we make use of the linear behavior of
the atom flux on the microwave output coupling power
(Fig.\,\ref{fig:det_detector}). We output couple a significant
number of atoms measurable by absorption imaging while still in the
weak output coupling regime. This number is compared to the number
of atoms detected by the cavity with the corresponding factor of the
output coupling powers.

We have calibrated the atom number in absorption imaging with the
atom number at the critical temperature which is well known for our
trap frequencies. For optimum settings of the cavity and laser
detunings we are able to detect $(24\pm5)\%$ of the output coupled
atoms with the cavity detector. This number is mainly limited by the
spatial overlap of the atom laser beam with cavity mode (see
Sec.\,\ref{sec:det_aiming}).

\subsection{\label{sec:det_aiming}Alignment of the atom laser beam with the cavity mode}
\begin{figure}[htb]
    \includegraphics[width=1\columnwidth, clip=true]{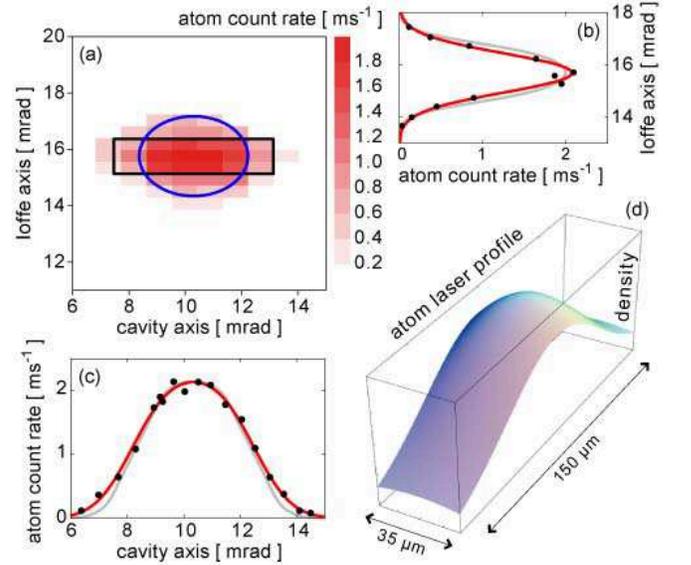}
    \caption{(a) The detected atom count rate for a constant atom flux is shown with respect to the inclination of the optical table along the two axes.
    The rectangle represents the active area of the cavity mode and the ellipse is the reconstructed size ($1/e$ diameter) of the atom
    laser at the position of the cavity.
    [(b) and (c)] Fit (red) to the measured data (black) by the convolution of the active size of the cavity mode with a Gaussian beam profile
    along the Ioffe (b) and cavity (c) axes. It is compared to the expected shape from numerical simulations of the Gross-Pitaevskii equation
    (gray).
    (d) Visualization of the extracted two-dimensional atom laser beam
    profile clipped by the active area of the cavity mode.}
    \label{fig:det_aiming}
\end{figure}

Obviously, in order to see single atoms with the cavity, the atom
laser has to propagate through the cavity mode. However, this was
not self-evident because the BEC production rig and the science
platform are completely independent entities of the experimental
apparatus and the alignment has to be better than a few millirad
without knowing the exact position of the cavity mode. Furthermore,
the second order Zeeman effect slightly bends the trajectory of the
atom laser in the \ket{F=2,m_{F}=0} state and modifies its final
lateral position by hundreds of micrometers. Although we have
aligned the cavity with respect to the BEC position as accurately as
possible with plummets during the assembly of the apparatus, the
atom laser did not innately hit the cavity mode. We correct these
deviations by tilting the whole optical table on which the
experiment rests employing its height adjustable legs. The tilt is
monitored with a dual-axis inclinometer\cite{inclinometer} having
its axes aligned along and perpendicular to the cavity axis. With
this method we aim the atom laser directly into the cavity mode
[Fig.\,\ref{fig:det_aiming}(a)] and maximize the atom count rate.

Moreover, tilting the experimental setup deterministically enables
us to deduce the diameter of the atom laser after a propagation of
36.4\,mm. The active area of the cavity mode is approximately
$35\times150\,\mum^2$. The size in the radial direction is
determined by the weakest detectable atom transits corresponding to
$\text{g}_0^{\text{min}} = 2\pi\,6.5$\,MHz. In the axial direction
it is given by the projection of the cavity length clipped by the
curved mirrors.

A deconvolution of the measured angle dependent count rates with
this active area, assuming a Gaussian atom laser beam profile,
yields $1/e$ diameters of 80 and 110\,\mum\ along and perpendicular
to the Ioffe axis, respectively [Figs.\,\ref{fig:det_aiming}(b) and
\ref{fig:det_aiming}(c), red]. The mapped atom laser, being output
coupled from the center of a Bose-Einstein condensate with
$1\times10^6$ atoms, is slightly inverted compared to the trap
geometry but almost round at the cavity. Here, its divergence along
the fast axis, i.e., cavity axis, is about 2\,mrad and less than
0.5\,mrad along the Ioffe axis, which makes it the best collimated
atom laser to date.\cite{bloch1999,hagley1999,le2001}

The repulsive mean field interaction from the remaining trapped BEC
is considerable only along the fast axis where it acts as a
defocusing lens for the atom laser beam. This results in an
expansion about four times larger than expected from Heisenberg's
uncertainty principle. Along the weakly confining Ioffe axis the
lensing effect is negligible and the size of the atom laser is
consistent with a free expansion of the initial ground state in the
trap. The atom laser size and therefore its divergence, especially
along the fast axis, can be further reduced by output coupling below
the center plane of the BEC\cite{le2001} and by smaller condensates
(see Sec.\,\ref{sec:inv_bec}).

We compare the measured atom laser profiles along its symmetry axes
with numerical simulations of the time evolution using the
Gross-Pitaevskii equation. The resulting density distributions of
the atom laser deviate slightly from a Gaussian
shape,\cite{kohl2005} but the measured convolutions with the cavity
mode agree very well with the simulated curves
[Figs.\,\ref{fig:det_aiming}(b) and \ref{fig:det_aiming}(c), gray].
The overestimated width along the Ioffe axis can be explained by the
angle of the BEC axis with respect to the horizontal plane, reducing
the spatial width of the output coupling region along the Ioffe
axis.

Along the cavity axis the slight deviation at the edges is probably
due to pointing variations, i.e., transverse oscillations of the
atom laser beam. Small collective oscillations in the trap are
translated into deflections of the atom laser beam over which we
integrate with our detection method. The collective oscillations,
mainly center of mass dipole oscillations in the trapped
Bose-Einstein condensate, can be excited by radio frequency
evaporation or incautious relaxation of the magnetic trap.

\begin{figure}[htb]
    \includegraphics[width=0.9\columnwidth, clip=true]{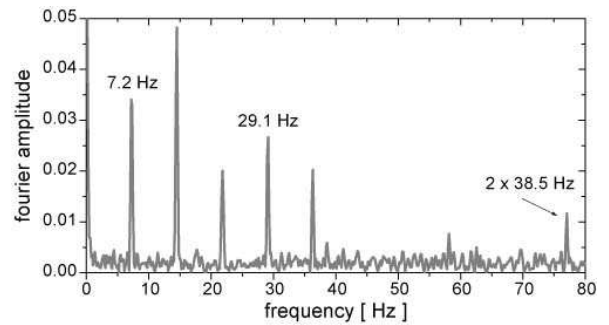}
    \caption{The Fourier spectrum of the detected atom laser flux
    exhibiting the trapping frequencies and their harmonics.
    A fast and precise tool to measure frequencies of collective oscillation in the trap.}
    \label{fig:det_oscillations}
\end{figure}

However, we can exploit this effect to precisely determine the
frequencies of excited collective oscillations in the trap by
analyzing the Fourier spectrum of the atom count rate. Such a
spectrum is shown in Fig.\,\ref{fig:det_oscillations} exhibiting
harmonics of the trap frequencies (dipole oscillations) and mutual
sidebands. The frequencies can be measured in situ with one and the
same experimental implementation of a Bose-Einstein condensate to
high precision (mHz), Fourier limited by the duration of the atom
laser recording.

\subsection{\label{sec:det_guiding}Guiding the atom laser}
The reason for the single atom detection efficiency not being unity
is mainly the mismatch of the atom laser and cavity mode sizes
[Fig.\,\ref{fig:det_aiming}(d)]. Their overlap is only about 50\%
assuming a box given by the projected length of the cavity mode and
a minimum peak atom field coupling strength of
$\text{g}_0^{\text{min}} = 2\pi\,6.5$\,MHz in the radial direction.
Calculating the atom trajectories taking into account the channeling
effect of dipole potential we find a maximum single atom detection
efficiency of 80\% and an averaged efficiency of about 50\% within
this box. This is in good agreement that we detect about one quarter
of the released atoms.

In order to increase the overlap and therefore the detectable number
of atoms it is possible to funnel the atoms with a dipole potential
created by a far red detuned guiding laser (850\,nm, 15\,mW, beam
waists of $30\times60\,\mum^2$) into the cavity mode. By doing so we
are able to improve the single atom detection efficiency by about a
factor of 2 to around 50\%. This number still differs from a perfect
detection efficiency because the dipole potential formed by the
probe laser is simply not strong enough to perfectly localize the
atoms in the axial direction at the antinodes of the standing wave.

Although we are able to increase the single atom detection
efficiency, employing the guiding laser involves some disadvantages.
The scattering and heating rate in the dipole potential formed by
the guiding laser can cause modifications of the atom arrival time
statistics which is undesirable for many experiments.\cite{ottl2005}
Furthermore, the guiding laser acts on both thermal and quantum
degenerate atoms and therefore diminishes a characteristic feature
of our detector, namely, the very sensitive discrimination of
thermal and condensed atom count rates (see
Sec.\,\ref{sec:investigation}).

%%%%%%%%%%%%%%%%%%%%%%%%%%%%%%%%%%%%%%%%%%%%%%%%%%%%%%%%%%%%%%%%%%%%%%%%%%%%%%%%% investigation

\section{\label{sec:investigation}\textsc{Investigation of Cold Atomic
Gases}}
The combination of a Bose-Einstein condensate with an ultrahigh
finesse optical cavity enables us to detect single atoms from a
quantum degenerate gas with very high sensitivity. Therefore we can
employ the cavity as a minimally invasive probe for cold atomic
clouds. This allows us to perform nondestructive measurements on the
ensemble of cold atoms \emph{in situ} and time resolved.

Assuming a constant weak output coupling power, the atom count rate
of the cavity detector depends on the properties of the source via
two factors. First of all the number of output coupled atoms is
proportional to the number of atoms fulfilling the resonance
condition, i.e., the one-dimensional density at the output coupling
plane. And secondly the atom count rate depends on the probability
for an output coupled atom to hit the cavity mode. Because of its
finite active area the cavity functions as a filter in momentum
space.

\subsection{\label{sec:inv_thermal}Thermal clouds}
\begin{figure}[t]
    \includegraphics[height=54mm, clip=true]{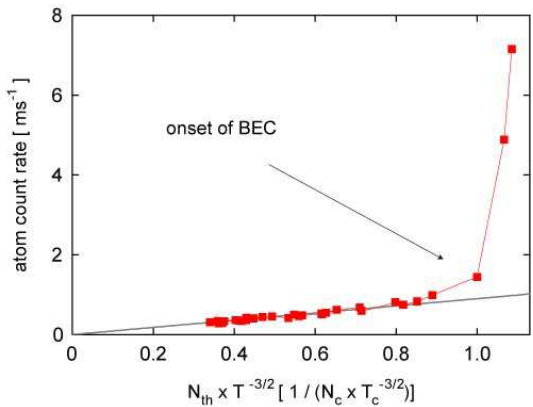}
    \caption{Investigation of atom count rates for thermal beams.
    The count rate is proportional to $N_{\text{th}}\,T^{-3/2}$ for temperatures above the critical temperature and
    sharply increases when cooling across the phase transition. Just above $T_{c}$ the density and momentum distributions of the thermal cloud
    are governed by a Bose distribution and obey a different scaling law as expected for a Gaussian distribution.}
    \label{fig:inv_thermal}
\end{figure}

For a thermal cloud the density at the output coupling central plane
is proportional to $N_{\text{th}}\,T^{-1/2}$ and the probability to
hit the detector is proportional to $1/T$ assuming Gaussian density
and momentum distributions. Therefore the thermal atom count rate
detected with the cavity is proportional to
$N_{\text{th}}\,T^{-3/2}$. This dependency is shown in
Fig.\,\ref{fig:inv_thermal}. At the critical temperature of $T_c
\approx 180$\,nK for $10^{7}$ atoms only about 0.6\% of the output
coupled thermal atoms will fly through the cavity mode and can
possibly be detected.

The onset of Bose-Einstein condensation can be clearly seen in the
sharp increase in the number of detected
atoms.\cite{seidelin2004,tychkov2006} Close to the critical
temperature, however, the detected atom flux slightly deviates from
the expected behavior because the approximated Gaussian
distributions for density and momentum are not valid anymore near
$T_{c}$. The thermal cloud is described by the more peaked Bose
distribution which yields an increased atom detection rate of about
30\% near the critical temperature of 180\,nK compared to the
Gaussian distribution.

\subsection{\label{sec:inv_bec}Quantum degenerate gases}
For Bose-Einstein condensates the probability for an atom to hit the
cavity mode and therefore the atom count rate detected with the
cavity is independent of temperature. The number of resonant atoms
participating in the output coupling process is proportional to the
density of the BEC and the area of the output coupling plane. This
means the atom flux is proportional to
$N_{\scriptscriptstyle{\text{BEC}}}^{4/5}$ because the Thomas-Fermi
radius of a BEC scales as
$N_{\scriptscriptstyle{\text{BEC}}}^{1/5}$.

\begin{figure}[t]
    \includegraphics[height=54mm, clip=true]{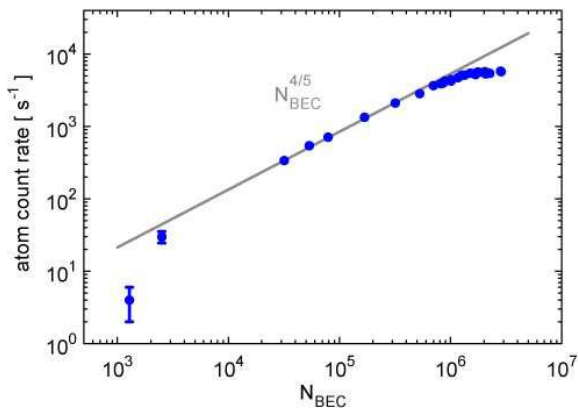}
    \caption{Investigation of detected atom count rates for
    pure quantum degenerate samples. The scaling with the atom number in a pure BEC exhibits three different regimes.
    The expected $N_{\scriptscriptstyle{\text{BEC}}}^{4/5}$ behavior is only valid for intermediate particle numbers.
    Very small and very large condensates obey different scaling laws due to an increased Heisenberg limited momentum spread and
    the mean field repulsion of the remaining condensate, respectively.}
    \label{fig:inv_bec}
\end{figure}

However, this dependency is only true for Bose-Einstein condensates
of intermediate size (Fig.\,\ref{fig:inv_bec}) and deviates for very
small and very large condensates. Output coupling from small
condensates accounts for a faster quantum mechanical expansion of
the initial ground state wave function in the atom laser. Therefore
the overlap between the transverse atom laser wave function and the
cavity mode is reduced. Large condensates on the other hand exhibit
increased divergence because of the mean-field repulsion exerted on
the atom laser propagating through the BEC. The condensate acts as
an imperfect diverging lens and displaces the maximum density
outward.\cite{le2001,kohl2005,riou2005} This results in a weaker
scaling of the detected atom flux with the number of atoms in the
BEC and possibly a decrease when the atom laser profile becomes more
``donut-mode-like.'' These three regimes are displayed in
Fig.\,\ref{fig:inv_bec} for measured atom count rates versus the
number of atoms in the ``pure'' BEC. The exact position of the
crossover between these regimes depends on the active area of the
single atom detector.

\subsection{\label{sec:inv_bimodal}Phase Transition}
Our single atom detector in form of the ultrahigh finesse optical
cavity is extremely sensitive and selective to quantum degenerate
atoms not only because of the increased density at the output
coupling region but also due to the filtering in transverse momentum
space.

\begin{figure}[hbt]
    \includegraphics[width=1\columnwidth, clip=true]{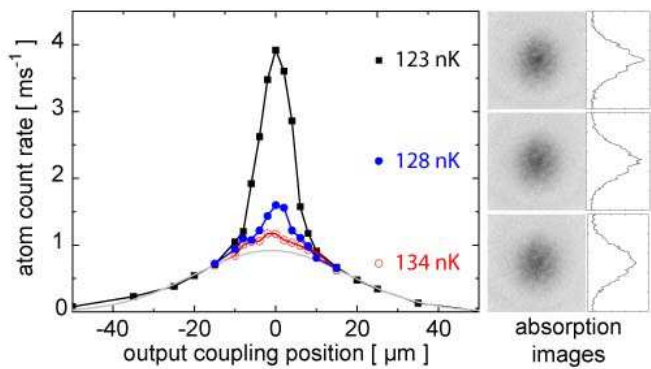}
    \caption{Analysis of the density distribution of the
    trapped ultracold atom gas by output coupling at different
    vertical positions relative to the center of the BEC and measuring the resulting atom count rate
    with the cavity. The profiles for three different temperatures
    around $T_{c}$ are shown in comparison with the absorption images.
    The high sensitivity of the cavity detector to quantum degenerate atoms allows for
    precise observation of the onset of Bose-Einstein condensation and the deviations from a Gaussian profile (gray curve).}
    \label{fig:inv_bimodal}
\end{figure}

This means we can more accurately observe the onset of Bose-Einstein
condensation as compared to absorption imaging techniques. The exact
determination of the critical temperature, in combination with the
precisely measured trap frequencies
(Fig.\,\ref{fig:det_oscillations}), allows one in turn to calibrate
the atom number obtained by the absorption images.

Furthermore we are able to survey the density distribution in the
trap along the vertical direction by scanning the resonant plane for
the output coupling process through the trapped cloud of cold atoms
(Fig.\,\ref{fig:inv_bimodal}). For temperatures close to the
critical temperature the density distribution of the thermal cloud
already deviates from the Gaussian shape and has to be described by
the more peaked Bose distribution [Fig.\,\ref{fig:inv_bimodal},
134\,nK]. For temperatures slightly below $T_{c}$ single atom
detection with the cavity allows us to observe and map very small
condensates that are not visible in absorption images
[Fig.\,\ref{fig:inv_bimodal}, 128\,nK and 123\,nK]. This could be a
valuable tool to study the temporal and spatial evolution of the
bosonic gas at the phase transition.

%%%%%%%%%%%%%%%%%%%%%%%%%%%%%%%%%%%%%%%%%%%%%%%%%%%%%%%%%%%%%%%%%%%%%%%%%%%%%%%%% conclusion

\section{\label{sec:conclusion}Discussion}
We have presented an apparatus that achieves the fusion of BEC
production with the single atom detection ability in the strong
coupling regime of cavity QED. The challenge to experimentally merge
these two fields was overcome by forging new paths for the
Bose-Einstein condensation setup and the ultrahigh finesse optical
cavity design.

The concept of the experimental realization is based on intertwined
technological modules: A nested vacuum system design with an
internal higher background pressure MOT chamber, an \emph{in vacuo}
magnetic transport arrangement cooled to $-90$\,\celsius and the
modular, exchangeable science platform providing access for samples
and probes to the Bose-Einstein condensate. The system is
distinguished by very reliable and reproducible operation for the
production of large Bose-Einstein condensates and stable, continuous
atom lasers. It features flexibility for research and applications
of atom lasers through the vast, free and accessible half-space
below the BEC.

In particular, we have implemented on the science platform a very
compact realization of an ultrahigh finesse optical cavity design
including a proper, UHV compatible vibration isolation system. With
this experimental setup we are able to detect single atoms from a
quantum degenerate source with high efficiency by aiming the atom
laser into the cavity mode. The atom laser allows for a very high
rate and controllable delivery of atoms into the cavity mode, which
facilitates research of cavity QED in the strong coupling regime
with single atoms. Moreover, the cavity as a single atom detector
opens up the field of quantum atom optics and is especially useful
to probe cold atomic clouds, particularly near the phase transition,
nondestructively \emph{in situ} and time resolved. Furthermore it is
an extremely sensitive tool to detect atomic beams for high
precision interferometry measurements and to investigate particle
correlations.\cite{ottl2005}

Future prospects with the system include single molecule detection
and the setup of a heterodyne detection technique\cite{mabuchi1999}
for probing the presence of an atom inside the cavity mode which
could be nondestructive on the atomic quantum state. Furthermore, we
intend to transport the Bose-Einstein condensate into the ultrahigh
finesse optical cavity. In the vertical direction we can apply a
moving optical standing wave formed by a far red detuned dipole
laser into which we could load the BEC and convey it downwards into
the cavity. The arrangement of coils around the cavity can be used
to apply magnetic field gradients for tomography experiments. The
lateral optical access enables us to create three-dimensional
optical lattices inside the ultrahigh finesse optical cavity which
will open the route to study strongly correlated systems with single
atom resolution.

\section*{\label{sec:acknowledgements}Acknowledgments}
We would like to thank Jean-Pierre Stucki, the mechanical workshop
of the ETH Physics Department, Thilo St\"oferle and Alexander
C.~Frank for aid in the construction and commissioning of the
apparatus, and Thomas Bourdel and Tobias Donner for experimental
assistance and valuable discussions. We acknowledge funding by SEP
Information Sciences, OLAQUI (EU FP6-511057), Qudedis (ESF), and
QSIT. Mention of industrial brand names is for technical
communication only and does not constitute an endorsement of such
products.

\end{document}